\RequirePackage{lineno}
\documentclass[pra,amsmath,amssymb,showpacs,notitlepage,nofootinbib,onecolumn]{revtex4-2}

\usepackage[utf8]{inputenc}

\usepackage{graphicx}
\usepackage[caption=false]{subfig}
\usepackage{xfrac}
\usepackage{amsmath}
\usepackage{float}
\usepackage{dsfont} 
\usepackage{physics} 
\usepackage{braket}
\usepackage{soul}
\usepackage{epsfig}
\usepackage{graphicx}
\usepackage{dcolumn}
\usepackage{bm}
\usepackage{amsmath}
\usepackage{amsthm}
\usepackage{amsfonts}
\usepackage{color} 
\usepackage{siunitx,booktabs}
\usepackage[hidelinks]{hyperref}
\usepackage{epstopdf}

\usepackage[english]{babel}
\usepackage{blindtext}

\allowdisplaybreaks 
\allowdisplaybreaks
\usepackage{amsmath}
\usepackage{amssymb}
\usepackage{amsfonts}
\usepackage{graphicx}

\usepackage{amsthm}

\let\oldequation\equation
\let\oldendequation\endequation

\renewenvironment{equation}
  {\linenomathNonumbers\oldequation}
  {\oldendequation\endlinenomath}

\let\oldeqnarray\eqnarray
\let\oldendeqnarray\endeqnarray

\renewenvironment{eqnarray}
  {\linenomathNonumbers\oldeqnarray}
  {\oldendeqnarray\endlinenomath}

\let\oldalign\align
\let\oldendalign\endalign

\renewenvironment{align}
  {\linenomathNonumbers\oldalign}
  {\oldendalign\endlinenomath}

\begin{document}
\title{Quantum key distribution with correlated sources}
\author{Margarida Pereira$^{1*}$} 
\author{Go Kato$^{2}$} 
\author{Akihiro Mizutani$^{3}$} 
\author{Marcos Curty$^{1}$}  
\author{Kiyoshi Tamaki$^{4*}$}
\affiliation{$^{1}$Escuela de Ingenier$\acute{\textit{\i}}$a de Telecomunicaci$\acute{o}$n, Department of Signal Theory and Communications, University of Vigo, Vigo E-36310, Spain \\
$^{2}$NTT Communication Science Laboratories, NTT Corporation, 3-1, Morinosato Wakamiya Atsugi-Shi, Kanagawa, 243-0198, Japan \\
$^{3}$ Mitsubishi Electric Corporation, Information Technology R\&D Center, 5-1-1 Ofuna, Kamakura-shi, Kanagawa, 247-8501 Japan\\
$^{4}$Faculty of Engineering, University of Toyama, Gofuku 3190, Toyama 930-8555, Japan \\
$^{*}$Corresponding authors: mpereira@com.uvigo.es, tamaki@eng.u-toyama.ac.jp}
%\date{\today}

%\makeatletter \AtEndDocument{\immediate\write\@auxout{\string\ulp@afterend}} 
%\AtBeginDocument{\immediate\openout\ulp@out=\jobname.upa\relax}
%
%\newcommand{\affvqcc}{Vigo Quantum Communication Center, University of Vigo, Vigo E-36315, Spain}
%\newcommand{\affuvigo}{Escuela de Ingeniería de Telecomunicación, Department of Signal Theory and Communications, University of Vigo, Vigo E-36310, Spain}
%\newcommand{\affatlantic}{atlanTTic Research Center, University of Vigo, Vigo E-36310, Spain}
%\newcommand{\afftoyama}{Faculty of Engineering, University of Toyama, Gofuku 3190, Toyama 930-8555, Japan}
%\newcommand{\affakihiro}{Mitsubishi Electric Corporation, Information Technology R\&D Center, 5-1-1 Ofuna, Kamakura-shi, Kanagawa, 247-8501 Japan}
%\newcommand{\affkatoOLD}{NTT Communication Science Laboratories, NTT Corporation, 3-1, Morinosato Wakamiya Atsugi-Shi, Kanagawa, 243-0198, Japan}
%
%\begin{document}
%	\title{Quantum key distribution with correlated sources}
%	\author{Margarida Pereira}	%\email{mpereira@com.uvigo.es}
%	\affiliation{\affuvigo}
%	 \author{Go Kato}
%    \affiliation{\affkatoOLD}
%    \author{Akihiro Mizutani}
%    \affiliation{\affakihiro}
%    \author{Marcos Curty}
%    \affiliation{\affuvigo}
%	\author{Kiyoshi Tamaki}  %\email{tamaki@eng.u-toyama.ac.jp}
%    \affiliation{\afftoyama}
%	\date{\today} 

\begin{abstract}
In theory, quantum key distribution (QKD) offers information-theoretic security. In practice, however, it does not due to the discrepancies between the assumptions used in the security proofs and the behaviour of the real apparatuses. Recent years have witnessed a tremendous effort to fill the gap, but the treatment of correlations among pulses has remained a major elusive problem. Here, we close this gap by introducing a simple yet general method to prove the security of QKD with arbitrarily long-range pulse correlations. Our method is compatible with those security proofs that accommodate all the other typical device imperfections, thus paving the way towards achieving implementation security in QKD with arbitrary flawed devices. Moreover, we introduce a new framework for security proofs, which we call the reference technique. This framework includes existing security proofs as special cases and it can be widely applied to a number of QKD protocols.
\end{abstract}

\maketitle

%%%%%%%%%%%%%%%%%%%%%%%%%%%%%%%%
\section{Introduction}
\label{sec:intro}
Quantum key distribution (QKD) allows two distant parties, Alice and Bob, to securely exchange cryptographic keys in the presence of an eavesdropper, Eve \cite{lo}. Despite the significant progress made in recent years, there is still a big gap between the information-theoretic security promised by the security proofs and the actual security offered by the practical implementations of QKD. The most pressing problem is the discrepancy between the idealised device models used in the security proofs and the functioning of the real devices employed in the experiments.  This is so because typical security proofs rely on assumptions to describe the behaviour of these devices and ignore their inherent imperfections. In practice, any deviation from these theoretical models might open security loopholes that could lead to side-channel attacks, thus compromising the security of QKD. A possible solution to this problem is to construct more realistic security proofs that can take into account device flaws. Indeed, lately, there have been notable advances in this direction. This includes, for example, the proposal of the decoy-state method \cite{hwang,lo3,wang2}, allowing the use of practical light sources while maintaining a high secret key rate. Also, measurement-device-independent QKD (MDI-QKD) \cite{lo2} can effectively eliminate all detector side channels, and is practical with current technology \cite{rubenok,silva,liu,tang,yin,comandar}. The missing step towards achieving implementation security in QKD is to better characterise and secure the parties' sources. 

Security loopholes in the source could emerge from three main causes: from state preparation flaws (SPFs) due to the finite precision of the modulation devices, from information leakage either due to side channels arising from mode dependencies or due to Trojan horse attacks (THAs) \cite{gisin2,vakhitov,lucamarini,tamaki3,wang}, or they could be caused by undesired classical correlations between the generated pulses. Mode dependencies of the emitted signals occur when the optical mode of a pulse depends on Alice's setting choices. That is, Alice's setting choices might be encoded in various degrees of freedom of the generated signals, not only on the desired one. Moreover, Eve can perform a THA by sending bright light into the source and then observe the back-reflected light to obtain partial information about Alice's internal settings. Finally, pulse correlations imply that the state of each pulse depends on the previous setting choices, such as  bit and basis choices. 

SPFs can be efficiently treated with the original loss-tolerant (LT) protocol \cite{tamaki}. This is so because in this scheme, the resulting secret key rate is almost independent of source's flaws. Its main drawback is the requirement that the states of the pulses are described by qubit states, which is hard to guarantee in practice due to unavoidable potential side channels. To address this limitation, a generalisation of the LT protocol was put forward very recently \cite{pereira}. This latter protocol encompasses SPFs, mode dependencies and THAs without requiring detailed information about the state of the side channels, which simplifies their experimental characterisation. There are also other techniques that can deal with mode dependencies and THAs, such as the Gottesman-Lo-L\"{u}tkenhaus-Preskill (GLLP) type security proofs involving the quantum coin idea \citep{gottesman,lo4,koashi2} (from now onwards, we shall refer to them as GLLP type security proofs) or the numerical approaches introduced in \cite{wang3,coles,winick}.

The final piece towards guaranteeing implementation security is, thus, to consider pulse correlations among the emitted signals. These pulse correlations are purely classical, and they arise from the limitations of practical modulators. In general, due to memory effects of these modulation devices, the state of a pulse depends not only on the actual modulation setting but also on the previous ones, meaning that the secret key information, i.e.~the bit and the basis choices, is encoded not only into a single pulse but also between subsequent pulses. Theoretically, it is believed that this correlation is very hard to model because the dimensionality of the state space becomes very large. In fact, all existing security proofs circumvent this imperfection by simply neglecting it, which means that they cannot guarantee the security of practical implementations. We remark that a few recent  works \cite{nagamatsu,mizutani,yoshino} have incorporated in their analysis certain pulse correlations between the emitted signals. However, all these works only consider restricted scenarios. In particular, the results in \cite{nagamatsu,mizutani} and in \cite{yoshino} only consider setting-choice-independent pulse correlations and intensity correlations between neighbouring pulses, respectively. Therefore, none of them can deal with pulse correlations in terms of the secret key information nor with long-range correlations. Another reason why these correlations have been ignored so far is because one expects that, in practice, they are small. Importantly, however, a small imperfection does not necessarily mean a small impact on the secret key rate, as Eve could in principle enhance such imperfection by exploiting, say, channel loss, resulting in a poor secret key rate \cite{gottesman,lo4,koashi2}. Therefore, we note that pulse correlations could be a serious threat to the security of QKD.

In this paper, we present a general and simple framework to guarantee the security of QKD in the presence of arbitrary classical pulse correlations of finite length. The key idea is very easy yet very useful, that is, we regard the leaked information encoded into the correlations of subsequent pulses as a side channel for each of the pulses. The key features of our method include: (1) when combined with the generalised loss-tolerant (GLT) protocol \cite{pereira} or with the reference technique (RT) introduced in this work, it can analytically guarantee the security of QKD with practical devices that suffer from typical source imperfections, i.e.~SPFs and side channels (including mode dependencies, THAs and pulse correlations), even if the state of the side channels is totally unknown; (2) due to its simplicity, our method is compatible with many other security proofs including those based on the inner product structure of the emitted pulses such as, for instance, the GLLP type security proofs \citep{gottesman,lo4,koashi2} and the numerical techniques in \cite{wang3,coles,winick}; and (3) our method can be applied to many QKD protocols such as, for example, the BB84 scheme \cite{bennett}, the six-state protocol \cite{bruss}, the SARG04 protocol \cite{scarani2}, distributed-phase-reference protocols \cite{inoue,takesue,stucki} and MDI-QKD \cite{lo2}. Our results indicate the feasibility of secure QKD with arbitrary flawed devices, and therefore they constitute an essential step towards closing the big gap between theory and practice in QKD. 

Also, a second contribution of this work is a new framework for security proofs, the RT, that can provide high performance in the presence of source imperfections. More precisely, this is a parameter estimation technique that includes existing security proofs as special cases (see the Supplementary Material). The RT incorporates the original LT protocol, and can reproduce the GLT protocol and the GLLP type security proofs. The key idea is to consider some reference states, which are close to the actual states prepared by the protocol of interest, and use them to simplify the estimation of the parameters needed to guarantee the security of the protocol. More precisely, by bounding the maximum deviation between the probabilities associated with the reference states and those associated with the actual states, one can obtain a relationship for the probabilities involving the actual states, based on those of the reference states. In doing so, one can estimate the parameters needed to guarantee the security of the actual protocol from the estimation that uses the reference states. We remark that the freedom to choose the reference states is very useful when dealing with source imperfections. In particular, this freedom allows us to analytically prove the security of a QKD protocol without any information on the side-channel states. This is important for achieving implementation security since a full characterisation of the side-channel states, which, in principle, could live in unknown physical modes, is certainly very challenging in practice. In this work, we consider three special cases of the RT and evaluate their secret key rate in the presence of pulse correlations and SPFs.

\section{Results}
\label{sec:results}
Pulse correlations occur, for instance, when the emitted signals depend on the previous values of the encoding device (e.g., a phase modulator). In other words, subsequent pulses leak information about Alice's former encoding choices. The key idea of our work to evaluate this complex scenario is to interpret these correlations as a side channel. By realistically modelling the source, we can bound this passive leakage of information and ensure secure QKD after performing enough privacy amplification. In what follows, we first outline the assumptions used in our security analysis, which is presented afterwards.

\subsection{Assumptions on Alice's and Bob's devices}
For simplicity, we consider a three-state protocol in which modulation devices are used to encode the bit and the basis choices. We do not explicitly consider the use of the decoy-state method \cite{hwang,lo3,wang2}; however, we remark that our framework could be combined with that method and also incorporate the effect of correlated intensity modulators as well as other imperfections of the intensity modulators \cite{tamaki3}. Furthermore, we assume an asymptotic scenario where Alice sends Bob an infinite number of pulses. We note, however, that the work presented here also applies to other protocols that employ more than three states, as discussed in the next section.

Additional assumptions might be required depending on the particular security proof technique that is combined with our method. For instance, if the RT based on the GLT protocol \cite{pereira} or the RT based on the original LT protocol \cite{tamaki}, which we will present below, are used, one also needs to assume that certain information about the states prepared by Alice is known. To be precise, for a setting choice $j\in\{0_Z,1_Z,0_X\}$ the state of the $k^{\rm th}$ pulse is in general purified into systems $C_kB_kE$ and expressed as
\begin{equation}
\ket{\psi_j}_{C_kB_kE} = a_j \ket{\phi_j}_{C_kB_k} \ket{{\lambda}}_{E} + \sqrt{1-{a_j}^2} \ket{\phi_j^\perp}_{C_kB_kE}.
\label{eq:main}
\end{equation}
Here, we take $a_j$ as a non-negative number satisfying $0 \le a_j \le 1$, which is possible by appropriately choosing the global phase of the states. The subscript $C_kB_kE$ stands for all the systems, which include not only the $k^{\text{th}}$ qubit (system $B_k$) that Alice sends to Bob over the quantum channel but also the system $C_k$, which is needed for purifying the state of system $B_k$, and $E$ is a system that includes Eve's system. System $E$ includes the systems sent by Alice over the quantum channel, such as, the back-reflected light from a possible THA, and the ancilla systems kept in Eve's lab. As we will discuss further later, in general, this system also includes Alice's ancilla systems employed in the virtual entanglement-based protocol, which is equivalent to the actual protocol. Some of the latter systems store the setting information for all the pulses sent before the $k^{\rm th}$ pulse. This means, in particular, that $\ket{\lambda}_{E}$ could depend on the setting choices for all the previous pulses. If it is not possible to find such a state, then $a_j$ becomes simply zero. From construction, Eq.~(\ref{eq:main}) is the most general state that can be prepared in a QKD protocol. In other words, Eq.~(\ref{eq:main}) simply decomposes a state $\ket{\psi_j}_{C_kB_kE}$ in a given Hilbert space into two states, each of which belongs to an orthogonal space. Precisely, one of them is the qubit state $\ket{\phi_j}_{C_k B_k}\ket{\lambda}_E$ (as the set of states $\{\ket{\phi_j}_{C_k B_k}\ket{\lambda}_E\}_j$ constitutes a qubit space), with $\ket{\lambda}_E$ being a state independent of the $k^{\rm th}$ setting choice, and the other is the setting-dependent side-channel state $\ket{\phi_j^\perp}_{C_k B_k E}$ that corresponds to unwanted and possibly unknown modes. This decomposition can always be done for an appropriate choice of $a_j$ with $0 \le a_j \le 1$. The characterisation of $\ket{\phi_j^\perp}_{C_kB_kE}$ is not required for the RT, and, in particular, no relationships between the states $\ket{\phi_j^\perp}_{C_kB_kE}$ and $\ket{\phi_{\tilde j}^\perp}_{C_kB_kE}$, and between $\ket{\phi_j}_{C_kB_kE}$ and $\ket{\phi_{\tilde j}^\perp}_{C_kB_kE}$ for $j \neq \tilde j$ are required, where $\tilde j$ represents a different setting choice to $j$. To use the RT, we only need to know a lower bound on the coefficient $a_j$ in Eq.~(\ref{eq:main}) and a full characterisation of the density operator of the qubit $B_k$. The main contribution of our work is to show that one can accommodate the effect of pulse correlations through the parameter $a_j$ in Eq. (\ref{eq:main}).

The assumptions on Bob's devices also depend on the security proof. For example, in the case of the RT based on the GLT protocol or based on the original LT protocol, one assumes that Bob measures the incoming pulses in the $Z$ or the $X$ basis. More precisely, Bob's measurements are represented by the positive-operator valued measures (POVMs) $\{\hat{{m}}_{0_Z},\hat{{m}}_{1_Z},\hat{{m}}_{f}\}$ and $\{\hat{{m}}_{0_X},\hat{{m}}_{1_X},\hat{{m}}_{f}\}$, respectively. Here, $\hat{{m}}_{\alpha\beta}$ corresponds to Bob obtaining the bit value $\alpha \in \{0,1\}$ when selecting the basis $\beta \in \{Z,X\}$, and $\hat{{m}}_{f}$ is associated with an inconclusive outcome. That is, we assume that these measurements satisfy the basis independent efficiency condition, i.e.~we impose that the operator $\hat{{m}}_{f}$ is the same for both basis. Note that, this condition is usually employed in security proofs to remove detector side-channel attacks exploiting channel loss \cite{lydersen,gerhardt}; however, it is not necessary in MDI-QKD, to which our framework also applies. Furthermore, we emphasise that our method to deal with pulse correlations could be used as well with security proofs where the basis independent efficiency condition is not guaranteed, such as in \cite{fung}.

\subsection{Security analysis in the presence of pulse correlations}
\label{sec:security}
In this section, we present the security analysis of QKD with pulse correlations. For this, we consider a security proof with the following properties. It employs an entanglement-based protocol where Alice prepares pulses in an entangled state, and she (Bob) measures the local (incoming) systems to distil a secret key. Also, it considers a particular detected pulse to estimate the phase error rate (or the phase error rate as a bound of the min-entropy). For simplicity, in what follows we shall explicitly mention only the phase error rate, but it applies to both cases. Security against coherent attacks can then be guaranteed with the help of Azuma's inequality \cite{azuma}, Kato's inequality \cite{kato} or by applying the techniques in \cite{christandl,dupuis}. Moreover, we assume that the security proof can be generalised such that it applies to a particular pulse with a side channel. That is, it can be used to prove the security of QKD in the presence of active and/or passive information leakage. Thanks to the reduction technique presented below, a particular pulse affected by correlations can be regarded as a pulse with a side channel, and therefore the security of QKD with pulse correlations is guaranteed. As an example, we now demonstrate that running a three-state protocol in the presence of nearest neighbour pulse correlations can be regarded as a three-state protocol in which each of the pulses entails side channels. We emphasise, however, that it is straightforward to generalise this reduction technique to an $m$-state protocol, as discussed below, and to arbitrarily long-range correlations (see the Materials and Methods section for more details).\\

\noindent \textit{Nearest neighbour pulse correlations} 

\noindent Let $\{\ket{\psi_j}_B\}_{j=0_Z,1_Z,0_X}$ be the set of three quantum states used in the three-state protocol. We assume that Alice chooses $\ket{\psi_j}_B$ with probability $p_j$ and sends the pulse prepared in the chosen state to Bob over the quantum channel. As for Bob's measurements, as already mentioned above, the assumptions vary according to the selected security proof. In an entanglement-based picture with nearest neighbour pulse correlations, the transmission of $n$ pulses by Alice can be described by first preparing $n$ ancilla systems $A$ and $n$ pulses in the state
\begin{equation}
\ket{\Psi}_{AB} = \sum_{j_1} \ket{j_1}_{A_1} \ket{\psi_{j_1}}_{B_1}  \sum_{j_2} \ket{j_2}_{A_2} \ket{\psi_{j_2|j_1}}_{B_2}  \hdots \sum_{j_n} \ket{j_n}_{A_n} \ket{\psi_{j_n|j_{n-1}}}_{B_n}, 
\label{eq:entanglement}
\end{equation}
and then by sending system $B$ to Bob. In Eq.~(\ref{eq:entanglement}),  $A = A_1, A_2, ..., A_n$ ($B = B_1, B_2, ..., B_n$) refers to the composite system of Alice's ancilla systems (Bob's pulses), where $A_k$ ($B_k$) for $k \in \{1,2,...,n\}$ denotes  Alice's $k^{\text{th}}$ ancilla system (Bob's $k^{\text{th}}$ pulse), the index $j_k\in\{0_Z,1_Z,0_X\}$, and $\{\ket{j_k}_{A_k}\}_{j_k\in\{0_Z,1_Z,0_X\}}$ is a set of unnormalised orthogonal states in a three dimensional Hilbert space with $||\ket{j_k}_{A_k}||=\sqrt{p_{j_k}}$, e.g., $||\ket{0_Z}_{A_k}||=\sqrt{p_{0_Z}}$. Importantly, $\ket{\psi_{j_k|j_{k-1}}}_{B_k}$ represents any nearest neighbour classical pulse correlation, namely, this is the state of the  $k^{\text{th}}$ emitted pulse when Alice selects the setting $j_k$, given that her previous setting choice was $j_{k-1}$.

Now, suppose that after Alice sends Bob system $B$, Bob obtains click events for some of the received signals. Then, Alice and Bob perform fictitious measurements on their systems to generate the raw data in the experiment in order. The secret key is distilled from the rounds in which they both perform $Z$ basis measurements. To prove the security of these rounds, we need to estimate the number of phase errors that Alice and Bob would have observed if they had performed their local measurements in a complementary basis instead. For this we consider that the users assign a tag $t \in \{0,\hdots,i\}$ to each round $k$ according to the value $t = k \mod (i+1)$, where $i$ is the maximum correlation length. In the case of nearest-neighbor pulse correlations we have that $t \in \{0,1\}$. The reason for this tag assignment will be explained later. Then, Alice and Bob construct $(i+1)$ virtual protocols and define their respective $t^{\rm th}$ sifted key as the subset of the total sifted key that originates from rounds with a tag $t$. In each of these virtual protocols, Alice and Bob perform $X$ basis measurements on the key rounds with tag $t$, and perform the same measurements as in the actual protocol on all the other rounds.

To estimate the phase error rate of the $t^{\rm th}$ virtual protocol, we now consider a particular pulse $k$ with a tag $t$. We are interested in the state of Alice's and Bob's $k^{\text{th}}$ systems before the fictitious measurements in the $t^{\rm th}$ virtual protocol, which resulted in a click at Bob's detectors. To obtain this state, recall that any operations and measurements on system $B$, including the detection measurements on the  pulses received by Bob, commute with Alice's measurements. Hence, we can assume that Alice has already measured her first $k-1$ ancillas before sending system B. Then, we have the resulting state as 
\begin{align}
&\ket{j_{1}'}_{A_1} \ket{\psi_{j_{1}'}}_{B_1} \hdots \ket{j_{k-1}'}_{A_{k-1}} \ket{\psi_{j_{k-1}'|j_{k-2}'}}_{B_{k-1}} \sum_{j_k}\ket{j_k}_{A_k} \ket{\psi_{j_k|j_{k-1}'}}_{B_k}  \nonumber \\
& ~~ \otimes \sum_{j_{k+1}}\ket{j_{k+1}}_{A_{k+1}} \ket{\psi_{j_{k+1}|j_{k}}}_{B_{k+1}} \hdots  \sum_{j_n} \ket{j_n}_{A_n} \ket{\psi_{j_n|j_{n-1}}}_{B_n},
\label{eq:protocol2}
\end{align}
where $j_{1}',\cdots, j_{k-1}'$ represent the outcomes of Alice's measurement on her first $k-1$ ancillas. Note that the state $\ket{\psi_{j_k|j'_{k-1}}}_{B_k}$ depends on the outcome $j'_{k-1} \in \{0_Z,1_Z,0_X\}$. If the round $k-1$ was a key round, then $j'_{k-1} \in \{0_Z,1_Z\}$. However, in the security proof we need to consider the case in which all key rounds are measured in the basis complementary to the $Z$ basis, leading to a contradiction. In order to resolve this issue, we proposed the aforementioned tagged virtual protocols. For example, in the $t^{\rm th}$ virtual protocol, the round $k$ is assigned a tag $t$ but $k-1$ is assigned another tag and therefore measured as in the actual protocol, meaning that $j'_{k-1} \in \{0_Z,1_Z,0_X\}$ as required. Note that the different virtual protocols could not have been run at the same time due to the non-commutativity of the $Z$ and $X$ basis measurements. However, the $t^{\rm th}$ virtual protocol allows us to prove the security of the $t^{\rm th}$ key and the security of the total key is ensured by the universal composability of each individual security proof. To simplify Eq.~(\ref{eq:protocol2}), we introduce the following definition
\begin{equation}
\ket{j_{k+1}}_{A_{k+1},\cdots, A_n, B_{k+2},\cdots, B_n}:=\ket{j_{k+1}}_{A_{k+1}} \sum_{j_{k+2}} \ket{j_{k+2}}_{A_{k+2}} \ket{\psi_{j_{k+2}|j_{k+1}}}_{B_{k+2}} \hdots \sum_{j_n} \ket{j_n}_{A_n} \ket{\psi_{j_n|j_{n-1}}}_{B_n}\,,
\label{eq:protocol21}
\end{equation}
which forms a set of orthogonal bases as $\{\ket{j_{k+1}}_{A_{k+1},\cdots, A_n, B_{k+2},\cdots, B_n}\}_{j_{k+1}=0_Z,1_Z,0_X}$. Also, we define the state
\begin{equation}
\ket{\lambda_{j_k}}_{A_{k+1},\cdots, A_n, B_{k+1},\cdots, B_n}:=\sum_{j_{k+1}}\ket{j_{k+1}}_{A_{k+1},\cdots, A_n, B_{k+2},\cdots, B_n,} \ket{\psi_{j_{k+1}|j_{k}}}_{B_{k+1}}\,.
\end{equation}
By using the above two states, we can rewrite Eq.~(\ref{eq:protocol2}) as
\begin{equation}
\ket{j_{1}'}_{A_1} \ket{\psi_{j_{1}'}}_{B_1}  \hdots \ket{j_{k-1}'}_{A_{k-1}} \ket{\psi_{j_{k-1}'|j_{k-2}'}}_{B_{k-1}} \sum_{j_k}\ket{j_k}_{A_k} \ket{\psi_{j_k|j_{k-1}'}}_{B_k} \ket{\lambda_{j_k}}_{A_{k+1},\cdots, A_n, B_{k+1},\cdots, B_n}\,.
\label{eq:protocol3}
\end{equation}
As a reference, recall that if there were no pulse correlations in the three-state protocol, the resulting state, instead of being in the form given by Eq.~(\ref{eq:protocol3}), would become 
\begin{equation}
\ket{j_{1}'}_{A_1} \ket{{\psi}_{j_{1}'}}_{B_1}  \hdots \ket{j_{k-1}'}_{A_{k-1}} \ket{{\psi}_{j_{k-1}'}}_{B_{k-1}} \sum_{j_k}\ket{j_k}_{A_k} \ket{{\psi}_{j_k}}_{B_k} \ket{\lambda}_{A_{k+1},\cdots, A_n, B_{k+1},\cdots, B_n},
\label{eq:protocol-no-correlation}
\end{equation}
where the state $\ket{\lambda}_{A_{k+1},\cdots, A_n, B_{k+1},\cdots, B_n}$ is independent of Alice's setting choice $j_k$, and can be expressed as
\begin{equation}
\ket{\lambda}_{A_{k+1},\cdots, A_n, B_{k+1},\cdots, B_n} = \sum_{j_{k+1}}\ket{j_{k+1}}_{A_{k+1}} \ket{{\psi}_{j_{k+1}}}_{B_{k+1}}  \hdots \sum_{j_n} \ket{j_n}_{A_n} \ket{{\psi}_{j_n}}_{B_n}.
\end{equation}

In the security proof for the three-state protocol without pulse correlations, one typically obtains the phase error rate by considering any attack on system $B_k$ in $\sum_{j_k}\ket{j_k}_{A_k} \ket{{\psi}_{j_k}}_{B_k}$ in Eq.~(\ref{eq:protocol-no-correlation}). On the other hand, when there are nearest neighbour pulse correlations, one can see from Eq.~(\ref{eq:protocol3}) that Alice's information $j_k$ is encoded not only on system $B_k$ but also on the systems $B_{k+1},\cdots, B_n$, and the state $\ket{\lambda_{j_k}}_{A_{k+1},\cdots, A_n, B_{k+1},\cdots, B_n,}$ serves as side-channel information about the state $\ket{\psi_{j_k|j_{k-1}'}}_{B_k}$. This suggests that, if we obtain the phase error rate for the composite systems $B_k$ and $B_{k+1},\cdots, B_n$ in  $\sum_{j_k}\ket{j_k}_{A_k} \ket{\psi_{j_k|j_{k-1}'}}_{B_k} \ket{\lambda_{j_k}}_{A_{k+1},\cdots, A_n, B_{k+1},\cdots, B_n}$, then the security follows. In other words, a protocol with pulse correlations can be simply regarded as a protocol where Alice prepares the states $\{\ket{\psi_{j_k|j_{k-1}'}}_{B_k} \ket{\lambda_{j_k}}_{A_{k+1},\cdots, A_n, B_{k+1},\cdots, B_n}\}_{j_k\in\{0_Z,1_Z,0_X\}}$ for {\it any} $k$ and sends systems $B_k$, $B_{k+1},\cdots, B_n$ to Bob.

Note that, our framework is also valid for the case where Alice emits mixed states instead of pure states. The emission of mixed states might happen due to imperfections in Alice's devices or when the prepared pure states are entangled with Eve's systems due to say a THA. To treat this latter scenario, the mixed states can be purified by introducing an ancilla system $C_k$, with $k \in \{1,2,\cdots,n\}$, which contains Alice's and Eve's systems. As a result, Eq.~(\ref{eq:protocol3}) becomes 
\begin{equation}
\ket{j_{1}'}_{A_1} \ket{\psi_{j_{1}'}}_{C_1 B_1}  \hdots \ket{j_{k-1}'}_{A_{k-1}} \ket{\psi_{j_{k-1}'|j_{k-2}'}}_{C_{k-1} B_{k-1}} \sum_{j_k}\ket{j_k}_{A_k} \ket{\psi_{j_k|j_{k-1}'}}_{C_k B_k} \ket{\lambda_{j_k}}_{A_{k+1},\cdots, A_n, C_{k+1} B_{k+1},\cdots, C_n B_n}.
\label{eq:protocol4}
\end{equation}
Again, if a security proof for the three-state protocol without pulse correlations shows that one can estimate the phase error rate for $\sum_{j_k} \ket{j_k}_{A_k} \ket{{\psi}_{j_k}}_{C_k B_k}$, it follows that $\sum_{j_k}\ket{j_k}_{A_k} \ket{\psi_{j_k|j_{k-1}'}}_{C_k B_k} \ket{\lambda_{j_k}}_{A_{k+1},\cdots, A_n, C_{k+1} B_{k+1},\cdots, C_n B_n}$ is also secure if one can obtain the parameters needed for the security proof given these latter states. Furthermore, we remark that, only for the purpose of estimating the phase error rate, in some cases it may make the mathematical analysis simpler to fictitiously consider an arbitrary attack on the systems $A_{k+1},\cdots, A_n$ (which, in reality, are inaccessible by Eve) besides the composite systems $B_k$ and $B_{k+1},\cdots, B_n$. Note that, the number of systems that we include as side channels does not matter, but what matters is how much the state $\ket{\lambda_{j_k}}_{A_{k+1},\cdots, A_n, B_{k+1},\cdots, B_n}$ depends on Alice's information $j_k$. Therefore, such fictitious attack on $A_{k+1},\cdots, A_n$ should not result in general in a lower key rate because these ancillas do not directly entail information about $j_k$. 

In either of these cases, once Alice and Bob obtain the phase error rate associated with the $t^{\rm th}$ virtual protocol, they can determine the amount of privacy amplification that they need to apply to the $t^{\rm th}$ sifted key to convert it into the $t^{\rm th}$ secret key. After doing so for all tags $t$, Alice and Bob define their final secret key pair as the concatenation of all the $t^{\rm th}$ secret keys.

Finally, we remark that all the discussions in this section and also in the next one do not require $j_k$ to be chosen from only three possibilities, i.e.~$\{0_Z,1_Z,0_X\}$. That is, by just considering $j_{k} \in \{1,2,3,\cdots,m \}$, our method applies for an $m$-state protocol.

\subsection{Particular device model}
Having stated the framework for the security proof in the presence of pulse correlations, we now consider a particular device model with only nearest neighbour pulse correlations. The purpose of this section is to show how to obtain the parameters needed in Eq.~(\ref{eq:main}) for a particular example of device model. Once this is achieved, one can directly apply the RT to guarantee the security of practical QKD implementations. We remark that, for simplicity, below we do not consider THAs or mode dependencies. However, they could readily be included by using the method in \cite{pereira}. Also, we assume that a single-photon source is available, and as a concrete example for modelling pulse correlations, we select the following instance of nearest neighbour pulse correlation 
\begin{equation}
\ket{\psi_{j_k|j_{k-1}'}}_{B_k}=\sqrt{1-\epsilon}\ket{{\phi}_{j_k}}_{B_k}+e^{i\theta_{j_k|j_{k-1}'}}\sqrt{\epsilon}\ket{{\phi}_{j_k}^{\perp}}_{B_k},
\label{particular nearest neighbor}
\end{equation}
for the three states. Here, $\ket{\psi_{j_k|j_{k-1}'}}_{B_k}$ is a single photon state living in a qubit space with $j_k \in \{0_Z,1_Z,0_X\}$, $\ket{{\phi}_{j_k}}_{B_k}$ is a qubit state, the parameter $\epsilon$ intuitively quantifies the strength of the correlation, $\theta_{j_k|j_{k-1}'}$ represents how the $k^{\text{th}}$ state depends on the previous information $j_{k-1}'$, and $\ket{{\phi}_{j_k}^{\perp}}_{B_k}$ is a state, in the same qubit space, that is orthogonal to $\ket{{\phi}_{j_k}}_{B_k}$.  Note that, when there are no pulse correlations, i.e.~$\epsilon=0$, the state $\ket{\psi_{j_k|j_{k-1}'}}_{B_k}$ becomes the  perfect state $\ket{\phi_{j_k}}_{B_k}$, which does not depend on the previous setting $j_{k-1}'$. However, in the presence of pulse correlations, i.e.~when $\epsilon > 0$, the overall state $\ket{\psi_{j_k|j_{k-1}'}}_{B_k}$ diverges from the ideal state $\ket{\phi_{j_k}}$, since it becomes dependent on the previous setting choice. The physical intuition of this model derives from the functioning of a phase modulator. To be precise, the state of an emitted pulse is typically affected by the modulation of the previous pulses such that there is a deviation depending on its pre-selected phase, which is quantified in the example given in Eq.~(\ref{particular nearest neighbor}) by $\theta_{j_k|j_{k-1}'}$. 

Below, we show how to derive the state in the form of Eq.~(\ref{eq:main}) for this particular example starting from Eq.~(\ref{particular nearest neighbor}). For this, we follow the idea introduced in the previous section and obtain the states $\ket{\psi_{j_k|j_{k-1}'}}_{B_k} \ket{\lambda_{j_k}}_{A_{k+1},\cdots, A_n, B_{k+1},\cdots, B_n}$ given by Eq.~(\ref{eq:protocol3}). By using Eq.~(\ref{particular nearest neighbor}), we have that
\begin{align}
&\ket{\psi_{j_k|j_{k-1}'}}_{B_k}\sum_{j_{k+1}}\ket{j_{k+1}}_{A_{k+1},\cdots, A_n, B_{k+2},\cdots, B_n} \ket{\psi_{j_{k+1}|j_{k}}}_{B_{k+1}} \nonumber \\
&=\left(\sqrt{1-\epsilon}\ket{{\phi}_{j_k}}_{B_k}+e^{i\theta_{j_k|j_{k-1}'}}\sqrt{\epsilon}\ket{{\phi}_{j_k}^{\perp}}_{B_k}\right) \nonumber \\
&\otimes\sum_{j_{k+1}}\ket{j_{k+1}}_{A_{k+1},\cdots, A_n, B_{k+2},\cdots, B_n} \left(\sqrt{1-\epsilon}\ket{{\phi}_{j_{k+1}}}_{B_{k+1}}+e^{i\theta_{j_{k+1}|j_{k}}}\sqrt{\epsilon}\ket{{\phi}_{j_{k+1}}^{\perp}}_{B_{k+1}}\right) \nonumber \\
&=:(1-\epsilon){\ket{\phi_{j_k}}_{A_{k+1},\cdots, A_n, B_k, B_{k+1},\cdots, B_n}}+\sqrt{1-(1-\epsilon)^2}\ket{{\phi}_{j_k|j_{k-1}'}^{\perp}}_{A_{k+1},\cdots, A_n, B_k, B_{k+1},\cdots, B_n},
\label{loss-tolerant-side-channel}
\end{align}
where 
\begin{equation}
{\ket{\phi_{j_k}}_{A_{k+1},\cdots, A_n, B_k, B_{k+1},\cdots, B_n}} = {\ket{\phi_{j_k}}_{B_k}} \sum_{j_{k+1}}\ket{j_{k+1}}_{A_{k+1},\cdots, A_n, B_{k+2},\cdots, B_n} \ket{{\phi}_{j_{k+1}}}_{B_{k+1}},
\end{equation}
\sloppy is a qubit state (note that, the set $\{\ket{\phi_{j_k}}_{A_{k+1},\cdots, A_n, B_k, B_{k+1},\cdots, B_n}\}_{j_k \in \{0_Z, 1_Z,0_X\}}$ spans a two-dimensional space) since $\sum_{j_{k+1}}\ket{j_{k+1}}_{A_{k+1},\cdots, A_n, B_{k+2},\cdots, B_n} \ket{{\phi}_{j_{k+1}}}_{B_{k+1}}$ is a normalised state independent of the information $j_k$, and $\ket{{\phi}_{j_k|j_{k-1}'}^{\perp}}_{A_{k+1},\cdots, A_n, B_k, B_{k+1},\cdots, B_n}$
is a state orthogonal to this qubit state. The explicit form of $\ket{{\phi}_{j_k|j_{k-1}'}^{\perp}}_{A_{k+1},\cdots, A_n, B_k, B_{k+1},\cdots, B_n}$ is omitted here for simplicity but it could be straightforwardly obtained from Eq.~(\ref{loss-tolerant-side-channel}). Importantly, we can regard our protocol as a protocol that uses the states in Eq.~(\ref{loss-tolerant-side-channel}) rather than the ideal states $\ket{{\phi}_{j_k}}_{B_k}$ for any $k$. We emphasise once again that the parameter $\epsilon$ and the state $\ket{{\phi}_{j_k|j_{k-1}'}^{\perp}}_{A_{k+1},\cdots, A_n, B_k, B_{k+1},\cdots, B_n}$ in Eq.~(\ref{loss-tolerant-side-channel}) represent most of the source imperfections (i.e.~SPFs, mode dependencies and THAs could be incorporated in a state of the form given by Eq.~(\ref{loss-tolerant-side-channel})) \cite{pereira}, not only pulse correlations. This comes from the generality of Eq.~(\ref{eq:main}).

Now, our formalism to deal with pulse correlations can be used directly with the RT, since the states in Eq.~(\ref{loss-tolerant-side-channel}) are in the form of Eq.~(\ref{eq:main}). For the RT (described in the next section) we only require to know a lower bound on the coefficient $1-\epsilon$ and a full characterisation of the state $\ket{\phi_{j_k}}_{B_k}$. We remark, however, that this framework can also be applied to  the numerical techniques in \cite{wang3,coles,winick} if, additionally, the form of the state $\ket{\phi_{j_k|j'_{k-1}}^{\perp}}_{A_{k+1},\cdots, A_n, B_k, B_{k+1},\cdots, B_n}$ is known, or if bounds involving the inner products ${}_{{A_{k+1},\cdots, A_n, B_k, B_{k+1},\cdots, B_n}}\bra{\phi_{j_k|j'_{k-1}}^{\perp}} \phi_{\tilde{j}_k|j'_{k-1}}^{\perp}\rangle_{A_{k+1},\cdots, A_n, B_k, B_{k+1},\cdots, B_n}$ and ${}_{{A_{k+1},\cdots, A_n, B_k, B_{k+1},\cdots, B_n}}\bra{\phi_{j_k}} \phi_{\tilde{j}_k|j'_{k-1}}^{\perp}\rangle_{A_{k+1},\cdots, A_n, B_k, B_{k+1},\cdots, B_n}$ for $j_k \neq \tilde{j}_k$ can be estimated, where $\tilde{j}_k$ represents a different setting choice to $j_k$.

Here, we restricted the discussion to the case of nearest neighbour pulse correlations, but our analysis also applies to arbitrarily long-range correlations, i.e. to any maximum correlation length $i$. For instance, these correlations could be characterised by  
\begin{eqnarray}
\big|{}_{B_k} \braket{{\psi}_{j_k|j_{k-1},\cdots,j_{w+1},\tilde{j}_w,j_{w-1},\cdots,j_1}|{\psi}_{j_k|j_{k-1},\cdots,j_{w+1},j_w,j_{w-1},\cdots,j_1}} {}_{B_k}\big|^2&\geq&1-\epsilon_{k-w},
\label{eq:assumption}
\end{eqnarray}
for any $w$ $(1 \leq w \leq n)$ and $k$ $(w+1 \leq k \leq \min\{n,w+i\})$. That is, the correlation could be characterised through the response according to the change of the $w^\text{th}$ index. In other words, we can quantify the correlation represented by $\epsilon_{k-w}$, where $k-w$ is the range of the correlation, by looking at the distinguishability of the states. Here, $i$ can be any non-negative number such that $i \ll n$, meaning that our method can incorporate arbitrary long-range correlations.
One can show that, from this model, it is straightforward to obtain the three states in the form given by Eq.~(\ref{eq:main}) (see the Materials and Methods section) and consequently apply the RT.

\subsection{Reference technique based on the original loss-tolerant protocol}
In this section, we introduce a new framework for security proofs, the RT, which results in a high secret key rate in the presence of source imperfections. In what follows, we outline the intuition behind the key idea of the RT by applying it to the original LT protocol \cite{tamaki}. A full description of the RT, including the detailed security proof, is presented in the Materials and Methods section. To simplify the discussion, here we shall assume collective attacks, however, our analysis can be generalised to coherent attacks (see the Materials and Methods section for more details). Just as an example, we consider a protocol with a single-photon source in the presence of side-channel information, such as pulse correlations, in which Alice prepares the following three states for each pulse emission 
\begin{equation}
{\ket{\psi_{j_k|j'_{k-1}}}_B = (1-\epsilon) \ket{\phi_{j_k}}_B + \sqrt{1-(1-\epsilon)^2} \ket{\phi_{j_k|j'_{k-1}}^\perp}_B,}
\label{eq:act_bef_new}
\end{equation}
where $B$ denotes the system to be sent to Bob. We remark that this subscript $B$ could be replaced with $A_{k+1},\cdots,A_n,B_k,B_{k+1},\cdots,B_n$ and then we would recover Eq.~(\ref{loss-tolerant-side-channel}). However, in this section we prefer to use Eq.~(\ref{eq:act_bef_new}) rather than Eq.~(\ref{loss-tolerant-side-channel}) for simplicity of notation. Note that, here we analyse the case of nearest neighbour pulse correlations, but the RT is also applicable to arbitrary long-range pulse correlations. In Eq.~(\ref{eq:act_bef_new}), $\ket{\phi_{j_k}}_B$ is a qubit state while $\ket{\phi_{j_k|j'_{k-1}}^\perp}_B$ corresponds to the side-channel state for $j_k \in \{0_Z,1_Z,0_X\}$ which lives in any dimensional Hilbert space and is orthogonal to $\ket{{\phi}_{j_k}}_B$ for each setting choice $j$. However, we do not assume any relationship between $\ket{\phi_{j_k|j_{k-1}'}}_B$ and $\ket{\phi_{\tilde j_{k}|j_{k-1}'}}_B$ for $j_k \neq \tilde j_{k}$. For instance, $\ket{{\phi}_{j_k}}_B$ can be defined as in \cite{pereira} such that
\begin{equation}
\begin{split}
&\ket{{\phi}_{0_Z}}_B = \ket{0_Z}_B, \\
&\ket{{\phi}_{1_Z}}_B = - \sin (\frac{\delta}{2})\ket{0_Z}_B +  \cos (\frac{\delta}{2}) \ket{1_Z}_B,\\ 
&\ket{{\phi}_{0_X}}_B = \cos (\frac{\pi}{4} + \frac{\delta}{4}) \ket{0_Z}_B +  \sin (\frac{\pi}{4} + \frac{\delta}{4}) \ket{1_Z}_B,
\end{split}
\label{eq:alice_states_new}
\end{equation}
where $\{\ket{0_Z}, \ket{1_Z}\}$ is a qubit basis and $\delta(\ge 0)$ is the deviation of the phase modulation from the intended value due to SPFs \cite{pereira}. That is, when there is no side-channel information, the states of the single-photons sent by Alice have the form given by Eq.~(\ref{eq:alice_states_new}), but in the presence of side-channel information, however, these states are defined by Eqs.~(\ref{eq:act_bef_new}) and (\ref{eq:alice_states_new}).

To prove the security of this QKD protocol with nearest-neighbour pulse correlations, we need to define two equivalent virtual protocols and estimate the phase error rate of each one of them individually, as explained in the previous section. In what follows, we consider a particular virtual protocol $t \in \{0,1\}$ and show how one can perform this estimation by using the RT. The key idea of the RT is to consider the phase error rate estimation that we would obtain if we replace the actual set of states of the protocol, $\{\ket{\psi_{0_Z|j'_{k-1}}}_B,\ket{\psi_{1_Z|j'_{k-1}}}_B,\ket{\psi_{0_X|j'_{k-1}}}_B\}$, with another set of states, which we call the reference states. Being the intuition that since the actual and the reference states are close to each other, one should be able to obtain a relationship between the events associated with the actual states by slightly modifying the relationship for the reference states. Note that, the choice of reference states is in principle infinite, however, for higher secret key rates they should be linearly dependent states such that unambiguous state discrimination (USD) \cite{chefles,dusek} is not possible. This allows us to use directly the original LT protocol \cite{tamaki} to estimate precisely some quantities associated with the reference states and their relationship, as an intermediate step towards obtaining the phase error rate associated with the actual states.

As an example, we select the reference states to be $\{\ket{\phi_{0_Z}}_B,\ket{\phi_{1_Z}}_B, \ket{\phi_{0_X}}_B\}$ that are defined in Eq.~(\ref{eq:alice_states_new}), and that  
correspond to the qubit part of the actual states in  Eq.~(\ref{eq:act_bef_new}). Also, we fictitiously consider that Alice chooses the reference states with the same probabilities as the actual states. Now, we can apply the RT in the following way. The first step is to find an expression for the probability of a phase error in terms of the reference states, which is a key parameter to be estimated in the security proof. For this, we consider an entanglement-based virtual protocol (see the Materials and Methods section for further details) employing the reference states, where Alice prepares the virtual states 
\begin{eqnarray}
\ket{\phi_{\alpha_X}^{\rm vir}}_B &=& \frac{\ket{\phi_{0_Z}}_B + (-1)^\alpha \ket{\phi_{1_Z}}_B }{\sqrt{2\left(1+(-1)^\alpha {}_{B}\bra{\phi_{0_Z}}\phi_{1_Z}\rangle_{B}\right) }},
\label{eq:vir_reference} 
\end{eqnarray}
with $\alpha \in \{0,1\}$ and where, for simplicity, we assumed that the selection probabilities in the $Z$ basis satisfy $p_{0_Z} = p_{1_Z}$. We can then define the probability of a phase error conditional on the reference states as 
\begin{eqnarray}
P({\rm ph|Ref}):=p_{1_X}^{\rm vir} p_{Z_B}{\rm Tr}\big[\dyad{\phi_{1_X}^{\rm vir}}{\phi_{1_X}^{\rm vir}}_B{\hat M}_{0_X}\big]+p_{0_X}^{\rm vir} p_{Z_B}{\rm Tr}\big[\dyad{\phi_{0_X}^{\rm vir}}{\phi_{0_X}^{\rm vir}}_B{\hat M}_{1_X}\big]\,,
\label{eq:ph_ref}
\end{eqnarray}
where $p_{\alpha_X}^{\rm vir}= \frac{1}{2}{p}_{Z_A}\big(1+(-1)^\alpha {}_{B}\bra{\phi_{0_Z}}\phi_{1_Z}\rangle_{B}\big)$ is the probability that Alice sends the virtual states defined in  Eq.~(\ref{eq:vir_reference}), $p_{Z_A}{:=p_{0_Z}+p_{1_Z}}$ ($p_{Z_B}$) is the probability that Alice (Bob) selects the $Z$ basis and ${\hat M}_{\alpha_X}$ is Bob's POVM element after any attack by Eve in the actual protocol. That is, ${\hat M}_{\alpha_X}:= \sum_{\tilde e} {\hat K}_{\tilde e} {\hat m}_{\alpha_X}{\hat K}_{\tilde e}^{\dagger}$ where ${\hat K}_{\tilde e}$ is the Kraus operator representing Eve's action in the actual protocol, ${\tilde e}$ corresponds to her measurement outcome, and ${\hat m}_{\alpha_X}$ is Bob's POVM element for detecting $\alpha_X$ in the actual protocol. The probabilities ${\rm Tr}\big[\dyad{\phi_{1_X}^{\rm vir}}{\phi_{1_X}^{\rm vir}}_B{\hat M}_{0_X}\big]$ and ${\rm Tr}\big[\dyad{\phi_{0_X}^{\rm vir}}{\phi_{0_X}^{\rm vir}}_B{\hat M}_{1_X}\big]$ in Eq.~(\ref{eq:ph_ref}) cannot be directly obtained since they involve reference and virtual states, which are never sent in reality. However, by exploiting the fact that the reference states are all qubit states, one can follow the idea of the original LT protocol \cite{tamaki} and get a simple relationship between these probabilities and the probabilities associated with the reference states. To see this, first note that in a qubit space the following expressions hold \begin{eqnarray}
&& \dyad{\phi_{1_X}^{\rm vir}}{\phi_{1_X}^{\rm vir}}_B = a \dyad{\phi_{0_Z}}{\phi_{0_Z}}_B + b \dyad{\phi_{1_Z}}{\phi_{1_Z}}_B - c \dyad{\phi_{0_X}}{\phi_{0_X}}_B,\nonumber \\
&& \dyad{\phi_{0_X}^{\rm vir}}{\phi_{0_X}^{\rm vir}}_B =  \dyad{\phi_{0_X}}{\phi_{0_X}}_B, \label{eq:qspace}
\end{eqnarray}
where the coefficients $a,b$ and $c$ are defined in the Materials and Methods section. We remark that, if there are no SPFs the coefficients become $a=b=c=1$. Then, by substituting Eq.~(\ref{eq:qspace}) into Eq.~(\ref{eq:ph_ref}) we obtain an expression for the probability of a phase error in terms of the reference states:
\begin{align}
0 &= p_{1_X}^{\rm vir} p_{Z_B}a{\rm Tr}\big[\dyad{\phi_{0_Z}}{\phi_{0_Z}}_B{\hat M}_{0_X}\big]+p_{1_X}^{\rm vir} p_{Z_B}b{\rm Tr}\big[\dyad{\phi_{1_Z}}{\phi_{1_Z}}_B{\hat M}_{0_X}\big]+p_{0_X}^{\rm vir} p_{Z_B}{\rm Tr}\big[\dyad{\phi_{0_X}}{\phi_{0_X}}_B{\hat M}_{1_X}]\nonumber\\
& -\Big[p_{1_X}^{\rm vir} p_{Z_B}c{\rm Tr}\big[\dyad{\phi_{0_X}}{\phi_{0_X}}_B{\hat M}_{0_X}\big]+ P(\rm ph|Ref)\Big].
\label{eq:eph1}
\end{align} In the RT, we call Eq.~(\ref{eq:eph1}) the Reference formula since it is used as a reference to obtain a similar expression in terms of the actual states. Note that, we cannot use the Reference formula directly in the security proof because it entails probabilities associated with the reference states, rather than the actual states.

Fortunately, by evaluating the deviation between the reference and the actual states, we can obtain bounds on the probabilities associated with the actual states and consequently the phase error rate. This part of the RT corresponds to the Deviation evaluation part (see the Materials and Methods section for further details). By following the analysis in the Supplementary Material, we have that this deviation is quantified by using
\begin{eqnarray}
g^L \Big( {\rm Tr}\big[\dyad{A}{A}{\hat M}\big],|\bra{A}{R}\rangle| \Big) \le {\rm Tr}\big[\dyad{R}{R}{\hat M}\big] \le g^U \Big( {\rm Tr}\big[\dyad{A}{A}{\hat M}\big],|\bra{A}{R}\rangle| \Big),
\label{eq:bound1}
\end{eqnarray}where $\hat{M}$ is any non-negative bounded operator such that $0 \le \hat{M} \le 1$ and, $\ket{A}$ and $\ket{R}$ are any normalised states associated with the actual and reference states respectively. To guarantee that $0 \leq \hat{M} \leq 1$, Alice and Bob must run the protocol sequentially, i.e.~Alice only emits the next pulse after Bob has measured the previous one. Here, the functions $g^L(x,y)$ and $g^U(x,y)$ are defined as 
\begin{align}
&g^L(x,y) = \left\{
        \begin{array}{ll}
            0 & \quad x  < 1 - y^2  \\
            x + (1-y^2)(1-2x) - 2y\sqrt{(1-y^2)x(1-x)} & \quad x  \geq 1 - y^2, 
        \end{array}
    \right.
\end{align}
and 
\begin{align}
g^U(x,y) = \left\{
        \begin{array}{ll}
            x + (1-y^2)(1-2x) + 2y\sqrt{(1-y^2)x(1-x)} & \quad x \leq y^2  \\
            1 & \quad x > y^2.
        \end{array}
    \right.
\end{align}Importantly, no measurement, including any measurement performed by Eve, can induce a larger deviation between the probabilities because Eq.~(\ref{eq:bound1}) holds for any ${\hat M}$. We remark that, here one could also use the trace distance argument \cite{tamaki3}, however, for the problem at hand, that bound is loose and, therefore, we employ a tighter bound. That is, we use the knowledge of the probability associated with the observable events in the actual protocol, i.e.~${\rm Tr}\big[\dyad{A}{A}{\hat M}\big]$, while the trace distance does not.

Now, we apply Eq.~(\ref{eq:bound1}) to the first and second lines of Eq.~(\ref{eq:eph1}) separately, thus converting Eq.~(\ref{eq:eph1}) into an expression for the probability of a phase error in terms of the actual states. For instance, note that the second line can be expressed by ${-} p_{Z_B} S_{-} \Tr \big [\dyad*{A_{-}}{A_{-}}_{CB} \hat{M}_{-} \big]$ with $\ket{A_{-}}_{CB}:=\sqrt{p_{1_X}^{\rm vir}c/S_{-}}\ket{0_{x,A}, X_{B}}_{C}\ket{\psi_{0_X|j'_{k-1}}}_{B}+\sqrt{p_{Z_A}/2S_{-}}\ket{0_{z,A}, Z_{B}}_{C}\ket{\psi_{0_Z|j'_{k-1}}}_{B}+\sqrt{{p_{Z_A}}/{2S_{-}}}\ket{1_{z,A}, Z_{B}}_{C}\ket{\psi_{1_Z|j'_{k-1}}}_{B}$ and ${\hat M}_{-}:=\hat P (\ket{0_{x,A}, X_{B}}_C) \otimes{\hat M}_{0_X} + \hat P \left([\ket{0_{z,A},Z_B}_C - \ket{1_{z,A},Z_B}_C]/\sqrt{2} \right)\otimes{\hat M}_{0_X}+\hat P \left([\ket{0_{z,A},Z_B}_C + \ket{1_{z,A},Z_B}_C]/\sqrt{2} \right) \otimes{\hat M}_{1_X},$ where $S_{-} = p_{1_X}^{\rm vir}c + {p}_{Z_A}$, system $C$ is an ancilla that stores the classical information associated with Alice's and Bob's setting choices, and $\hat P (\ket{\cdot}) = \dyad{\cdot}{\cdot}$ (see the Materials and Methods section). Importantly, here we have  mathematically represented the summed probabilities using the trace. By obtaining a similar expression for the first line of Eq.~(\ref{eq:eph1}), we find that this equation becomes (see the Materials and Methods section)
\begin{align}
0&\le S_{+} {g^U} \bigg(\frac{p_{1_X}^{\rm vir}p_{Z_B}a}{S_{+}} \Tr\big[\dyad*{\psi_{0_Z|j'_{k-1}}}{\psi_{0_Z|j'_{k-1}}}_B{\hat M}_{0_X}\big]+ \frac{p_{1_X}^{\rm vir}p_{Z_B}b}{S_{+}} \Tr \big[\dyad*{\psi_{1_Z|j'_{k-1}}}{\psi_{1_Z|j'_{k-1}}}_B {\hat M}_{0_X}\big] \nonumber \\
&+\frac{p_{0_X}^{\rm vir}p_{Z_B}}{S_{+}}\Tr \big[\dyad*{\psi_{0_X|j'_{k-1}}}{\psi_{0_X|j'_{k-1}}}_B{\hat M}_{1_X}\big],1-\epsilon \bigg) - S_{-} {g^L}\bigg(\frac{p_{1_X}^{\rm vir}p_{Z_B}c}{S_{-}} \Tr \big[\dyad*{\psi_{0_X|j'_{k-1}}}{\psi_{0_X|j'_{k-1}}}_B{\hat M}_{0_X}\big] \nonumber \\ 
&+ \frac{P(\rm ph|Act)}{S_-},1-\epsilon \bigg),
\label{eq:eph2}
\end{align}
where $S_+ = p_{1_X}^{\rm vir}a + p_{1_X}^{\rm vir}b + p_{0_X}^{\rm vir}$, and 
\begin{eqnarray}
P({\rm ph|Act}):={\tilde p}_{1_X}^{\rm vir} p_{Z_B}{\rm Tr}\big[\dyad*{\psi_{1_X|j'_{k-1}}^{\rm vir}}{\psi_{1_X|j'_{k-1}}^{\rm vir}}_B {\hat M}_{0_X}\big]+{\tilde p}_{0_X}^{\rm vir} p_{Z_B}{\rm Tr}\big[\dyad*{\psi_{0_X|j'_{k-1}}^{\rm vir}}{\psi_{0_X|j'_{k-1}}^{\rm vir}}_B{\hat M}_{1_X}\big],
\label{eq:phase_errors}
\end{eqnarray} 
is the probability of a phase error conditional on the actual states. In Eq.~(\ref{eq:phase_errors}), $\ket{\psi_{\alpha_X|j'_{k-1}}^{\rm vir}}_B$ are the virtual states associated with the actual states and ${\tilde p}_{\alpha_X}^{\rm vir}$ are their respective probabilities. The explicit form of $\ket{\psi_{\alpha_X|j'_{k-1}}^{\rm vir}}_B$ is omitted here for simplicity, but similarly to Eq.~(\ref{eq:vir_reference}), $\ket{\psi_{\alpha_X|j'_{k-1}}^{\rm vir}}_B \propto \ket{\psi_{0_Z|j'_{k-1}}}_B + (-1)^\alpha \ket{\psi_{1_Z|j'_{k-1}}}_B$. Importantly, Eq.~(\ref{eq:eph2}) is valid for any eavesdropping strategy, {i.e.~any Kraus operator ${\hat K}_{\tilde e}$,} that is included in the operators ${\hat M_{\alpha_X}}$ (see the discussion just after Eq.~(\ref{eq:ph_ref})), and it can be directly used for the phase error estimation. To clearly see how Eq.~(\ref{eq:eph2}) is related with quantities observed in an actual experiment, we write 
\begin{eqnarray}
0&\le& S_{+} {g^U}\bigg(\frac{p_{1_X}^{\rm vir}p_{Z_B}a}{S_{+}{p_{0_Z}p_{X_B}}} {P(q_{0z,0x}|{\rm Act})}+  \frac{p_{1_X}^{\rm vir}p_{Z_B}b}{S_{+}{p_{1_Z}p_{X_B}}} {P(q_{1z,0x}|{\rm Act})} +\frac{p_{0_X}^{\rm vir}p_{Z_B}}{S_{+}{p_{0_X}p_{X_B}}} {P(q_{0x,1x}|{\rm Act})},1-\epsilon \bigg) \nonumber\\
&&- S_{-} {g^L}\bigg(\frac{p_{1_X}^{\rm vir}p_{Z_B}c}{S_{-}{p_{0_X}p_{X_B}}} P(q_{0x,0x}|{\rm Act})+ \frac{P(\rm ph|Act)}{S_-},1-\epsilon \bigg),
\label{eq:eph3}
\end{eqnarray}
where, for example, ${P(q_{0z,0x}|{\rm Act})} := {p_{0_Z} p_{X_B}} \Tr \big[\dyad*{\psi_{0_Z|j'_{k-1}}}{\psi_{0_Z|j'_{k-1}}}_B \hat{M}_{0_X}\big]$ is the joint probability (i.e.~the yield) that Alice selects the setting $0_Z$ and prepares the state $\ket{\psi_{0_Z|j'_{k-1}}}_B$, and Bob's measurement outcome is $0_X$. Finally, by solving Eq.~(\ref{eq:eph3}) with respect to $P({\rm ph|Act})$ we obtain the probability of a phase error. The phase error rate of the $t^{\rm th}$ sifted key is then defined as $e_{X} = P({\rm ph|{\rm Act}})/Y_Z$, where $Y_Z:= P(q_{0z,0z}|{\rm Act})+P(q_{0z,1z}|{\rm Act})+P(q_{1z,0z}|{\rm Act})+P(q_{1z,1z}|{\rm Act})$ is the yield in the $Z$ basis, i.e.~the joint probability that Alice and Bob choose the $Z$ basis and Bob obtains a detection event.

\subsection{Simulation of the secret key rate}
To show the performance of QKD in the presence of pulse correlations we now present the simulation results. For simplicity of discussion, here 
we apply our framework to two different cases of the RT: the RT based on the GLT protocol \cite{pereira} and the RT described in the previous section. We remark that, the GLLP type security proofs \cite{gottesman,lo4,koashi2} are also regarded as a special case of the RT where we select the actual states as the reference states and skip the Reference formula part (see the Supplementary Material for the proof of this claim). However, they involve four states, rather than three states, and analytical or numerical optimisation is required. The comparison between the RT based on the GLLP type security proofs and the RT based on the original LT protocol is presented in the Supplementary Material.

The main difference between the RT based on the GLT protocol and the RT based on the original LT protocol is that, in the former, a different bound is employed to estimate the probabilities associated with the actual states. More precisely, the RT based on the GLT protocol essentially uses an inequality involving eigenvalues, instead of Eq.~(\ref{eq:bound1}), which has the form  
\begin{eqnarray}
\Tr \big[\dyad{\phi_{j_k}}{\phi_{j_k}}_B \hat{M}_{\alpha_X}\big] + \lambda_{j_k}^{\rm min} \le \Tr \big[\dyad*{\psi_{0_Z|j'_{k-1}}}{\psi_{0_Z|j'_{k-1}}}_B \hat{M}_{\alpha_X}\big] \le \Tr \big[\dyad{\phi_{j_k}}{\phi_{j_k}}_B \hat{M}_{\alpha_X}\big] + \lambda_{j_k}^{\rm max}.
\label{eq:GLTcoef}
\end{eqnarray} 
Here, $\lambda^{\rm min}_{j_k}$ and $\lambda^{\rm max}_{j_k}$ are the eigenvalues of a matrix in the form $\Big[\begin{smallmatrix} C_{j_k}&B_{j_k}^*\\ B_{j_k}&0 \end{smallmatrix}\Big]$, where $B_{j_k} = (1-\epsilon)\sqrt{1-(1-\epsilon)^2}$ and $C_{j_k} = 1-(1-\epsilon)^2$. Importantly, the inequality in Eq.~(\ref{eq:GLTcoef}) is valid for any ${\hat M}_{\alpha X}$ with $\alpha\in \{0,1\}$, and therefore we can use it to consider the deviation between the probabilities associated with the reference states and the ones associated with the actual states (see \cite{pereira} for more details). We emphasise that pulse correlations are not taken into account in \cite{pereira}, however, we can apply our method to deal with pulse correlations to this security analysis. In doing so, we simply consider a QKD protocol with the states in Eqs.~(\ref{eq:act_bef_new}) and (\ref{eq:alice_states_new}), and apply the RT based on the GLT protocol. That is, besides pulse correlations we also include the effect of SPFs by assuming $\delta>0$ in Eq.~(\ref{eq:alice_states_new}). Furthermore, recall that system $B$ in Eqs.~(\ref{eq:act_bef_new}), (\ref{eq:alice_states_new}) and (\ref{eq:GLTcoef}) can include more systems, not only those sent to Bob. Indeed, in these equations the subscript $B$ could be replaced by $ {A_{k+1},\cdots, A_n, B_k, B_{k+1},\cdots, B_n}$, allowing us to consider pulse correlations as the side channel. Note that, to simplify the mathematical analysis we do not trace out Alice's subsequent systems $A_{k+1},\cdots,A_n$.
Since these systems are independent of the setting $j_k$, they do not provide any relevant information to Eve and therefore they do not affect our estimation of the phase error rate.

For the simulations, we assume the asymptotic regime where the secret key rate formula for a single-photon source can be expressed as
\begin{equation}
R \ge Y_Z \left(1 - h(e_X) - fh(e_Z)\right),
\label{eq:key_rate_R}
\end{equation}
where, as defined before, $Y_Z$ is the yield in the $Z$ basis and $e_X$ is the phase error rate. The term $e_Z$ is the bit error rate, $h(x) = -x\log_2 (x) - (1-x)\log_2 (1-x)$ is the binary entropy function, and $f$ is the error correction efficiency. We remark that Eq.~(\ref{eq:key_rate_R}) refers to the secret key rate obtainable per emitted $t$-tagged pulse. However, in practice, disregarding finite-size effects, one expects the observed data to be the same independently of the tag $t$. Therefore, for simplicity, in the simulations we assume that Eq.~(\ref{eq:key_rate_R}) represents the overall secret key rate obtainable per emitted pulse. Note that, $Y_Z$ and $e_Z$ are directly observed in a practical implementation of the protocol, but in the simulations a channel model (see \cite{pereira} for more details) is employed instead. The experimental parameters used are: dark count rate
of Bob's detectors $p_d=10^{-7}$, $f = 1.16$ and the probabilities for Alice and Bob to select the $Z$ basis are, for simplicity, ${p}_{Z_A} = \frac{2}{3}$ and ${p}_{Z_B} = \frac{1}{2}$. Unfortunately, there are no quantitative works characterising pulse correlations  (i.e.~the value of the parameter $\epsilon$) therefore, for illustration purposes, we select the values $10^{-3}$ and $10^{-6}$ to evaluate this imperfection. Also, in order to investigate how the length of the pulse correlations affects the secret key rate, we consider the nearest neighbour correlation $\epsilon_1$, as well as correlations among two subsequent pulses, $\epsilon_2$, and among ten subsequent pulses, $\epsilon_{10}$ (see Eq.~(\ref{eq:assumption}) for the definitions of these epsilon parameters). Regarding SPFs, we choose $\delta = 0$ and $\delta = 0.063$ according to the experimental results reported in \cite{xu2,honjo,li}. The results for the RT based on the GLT protocol and for the RT based on the original LT protocol are illustrated in Fig.~1.

\begin{figure}[h!]
	\centering
	 \subfloat[]{\includegraphics[width=0.4\columnwidth]{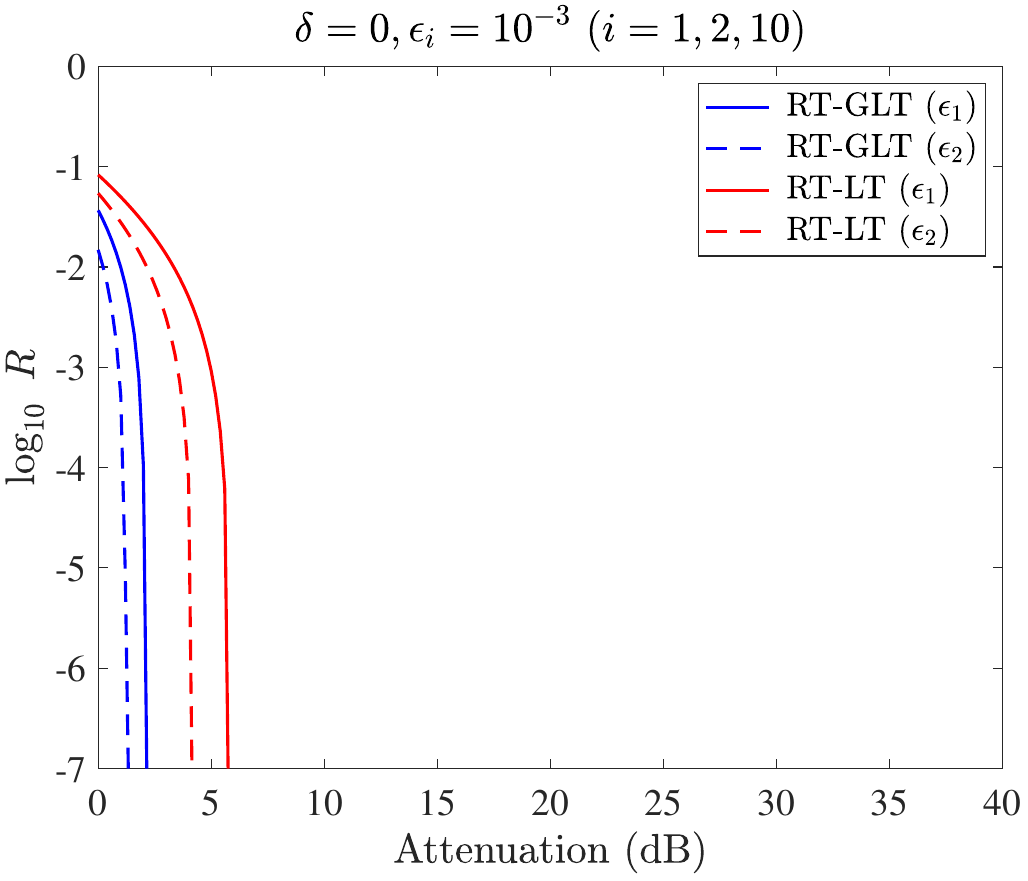}}
     ~~~
    \subfloat[]{\includegraphics[width=0.4\columnwidth]{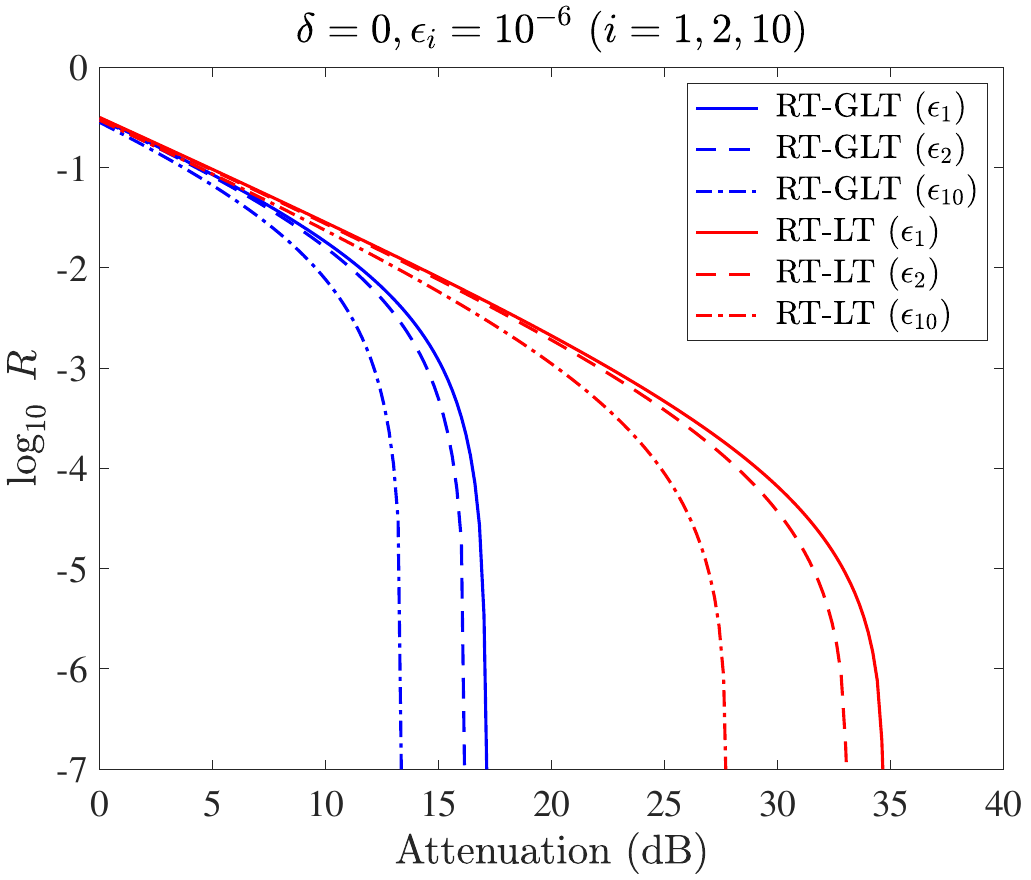}} \\
  	\subfloat[]{\includegraphics[width=0.4\columnwidth]{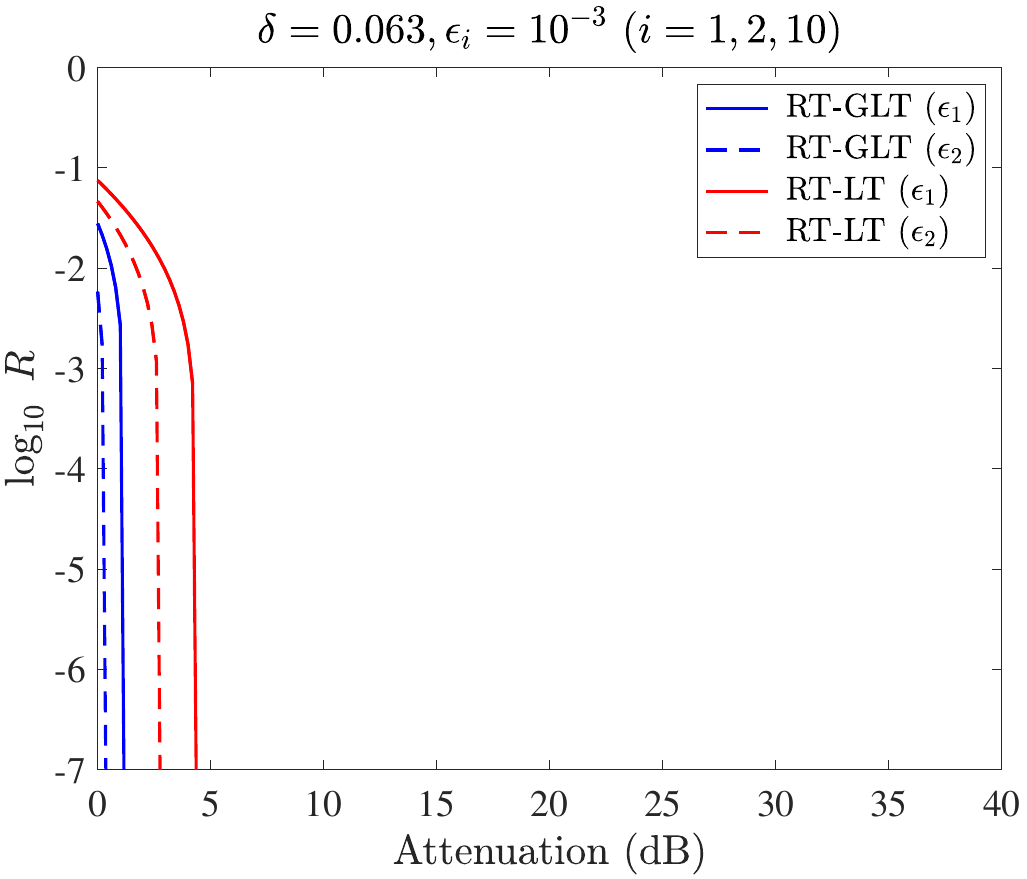}}
     ~~~
    \subfloat[]{\includegraphics[width=0.4\columnwidth]{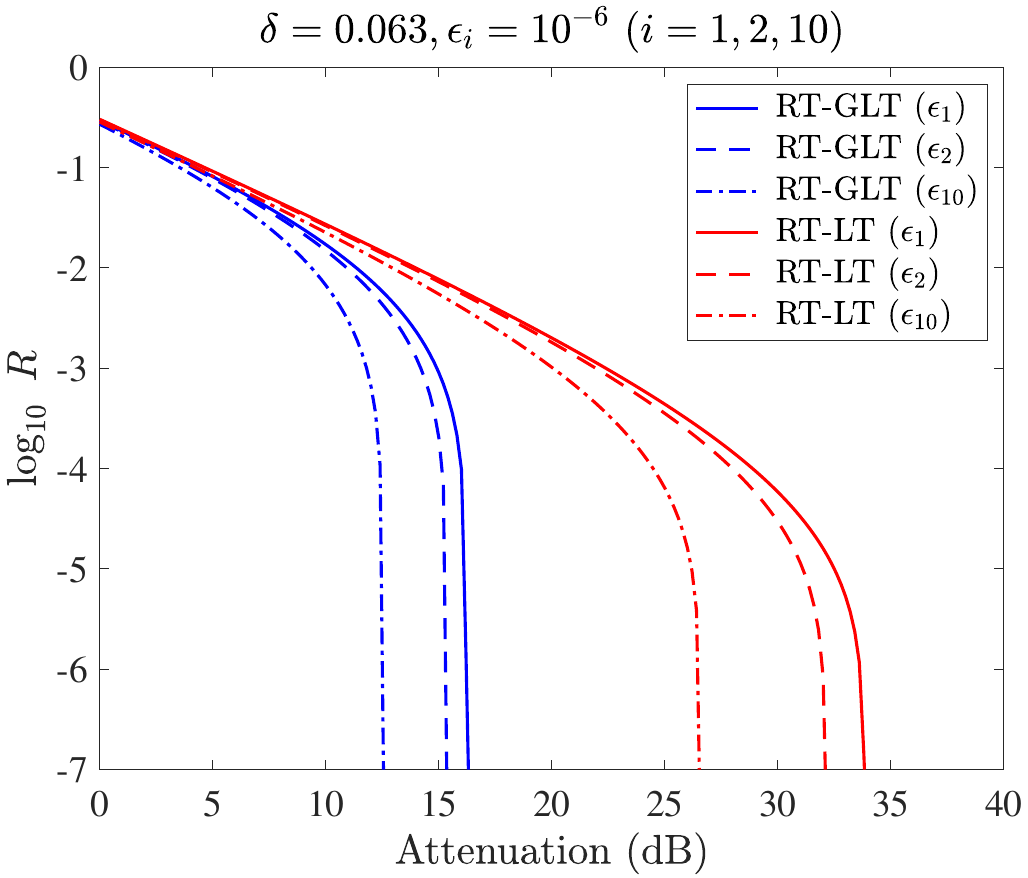}}
	\caption{Secret key rate $R$ against the overall system loss measured in dB in the presence of correlations between the emitted pulses when our method is applied to the reference technique based on the generalised loss-tolerant (RT-GLT) protocol and to the reference technique based on the original loss-tolerant (RT-LT) protocol. In all graphs, the blue and red lines are associated with the RT-GLT and the RT-LT, respectively. The solid lines correspond to the nearest neighbour pulse correlations $\epsilon_1$, while the dashed (dashed-dotted) lines correspond to second $\epsilon_2$ (tenth $\epsilon_{10}$) neighbour pulse correlations as indicated in the legend. (a) When there are no SPFs and the parameter $\epsilon$ is high, the RT-GLT and the RT-LT provide similar secret key rates. (b) As the parameter $\epsilon$ decreases, both security proofs provide higher secret key rates but the RT-LT clearly outperforms the RT-GLT. (c) In the presence of SPFs the secret key rate is only slightly worse for all cases, since the security proofs are based on the LT protocol. (d) For high SPFs and low $\epsilon$, the RT-LT is still superior to the RT-GLT.}
	\label{fig:comparison1}
\end{figure}

As expected, this figure shows that when the magnitude of pulse correlations characterised by $\epsilon_i$ increases the secret key rate decreases. Also, as the length of the correlations taken into account increases the secret key rate drops. We note, however, that even when long-range correlations are considered, a secret key can still be obtained. Namely, Fig.~1 shows that, for $\epsilon = 10^{-6}$, one can generate a secret key even when there are correlations between ten subsequent pulses. Clearly, for a smaller value of the parameter $\epsilon_i$ longer correlations can be included. In fact, if $\epsilon_i$ is small enough, one can consider a very long range of pulse correlations while guaranteeing the security of QKD. 

We emphasise that, the security proof selected highly affects the results obtained, and this is also illustrated in Fig.~1, where we apply our technique to two different cases of the RT. To compare the RT based on the GLT protocol and the RT based on the original LT protocol as a function of pulse correlations one can examine Figs.~1(a) and 1(b), or Figs.~1(c) and 1(d). Noticeably, as the magnitude of the pulse correlation $\epsilon_i$ increases, the secret key rate deteriorates for both of them. However, the RT based on the LT protocol outperforms the RT based on the GLT protocol in all the parameter regimes investigated. Also, by comparing Figs.~1(a) and 1(c), or Figs.~1(b) and 1(d), one can see the effect of SPFs. As expected, the RT based on the GLT protocol and the RT based on the LT protocol are barely affected by this imperfection since they inherit, from the GLT protocol and the original LT protocol respectively, high tolerance against SPFs with channel loss. The big difference observed in Fig.~1 between these two cases of the RT arises because of the following reason. Recall that we need to evaluate the deviation between the probabilities associated with the reference states and those associated with the actual states. For this, the bound employed in the RT based on the GLT protocol is obtained by calculating certain eigenvalues and thus they entail square root terms, which deteriorate the secret key rate. Note that, in the trace distance argument \cite{tamaki3} square root terms are also present, resulting in loose bounds. On the other hand, the RT based on the original LT provides a tighter estimation of the phase error rate thanks to the bound in Eq.~(\ref{eq:bound1}). More precisely, the square root terms in Eq.~(\ref{eq:bound1}) include detection probabilities, which decrease as the channel loss increases, while for the other two bounds the square root terms are constant, and thus the high performance is maintained by using the bound in Eq.~(\ref{eq:bound1}). Finally, we remark again that, the RT framework is general and can be applied to other QKD protocols as well, as shown in the Supplementary Material.

\section{Discussion}
\label{sec:discussion}
Security proofs of QKD have to consider source imperfections in the theoretical models. Fortunately, state preparation flaws (SPFs), Trojan horse attacks  \cite{gisin2,vakhitov,lucamarini,tamaki3,wang} and mode dependencies have been considered together very recently in \cite{pereira}. In this work, we have introduced a general framework to deal with pulse correlations, which are the last piece required for securing the source. Importantly, our framework is compatible with those security proofs that incorporate other source imperfections, and therefore it can be used to guarantee implementation security with flawed devices, by combining it with measurement-device-independent QKD \cite{lo2} and the results in \cite{pereira}. We remark that the decoy-state method \cite{hwang,lo3,wang2} has not been considered in this work, and, therefore, the imperfections of the intensity modulator have not been addressed. However, these imperfections could be straightforwardly  included in our framework. The key idea for dealing with pulse correlations is interpreting the information encoded in the subsequent pulses as side-channel information. By doing so, we have shown that, as long as the magnitude of the correlations is small, a secret key can still be obtained even when there are correlations over a long range of pulses. Moreover, our framework can be directly applied in combination with existing security proofs such as the generalised loss-tolerant (GLT) protocol \cite{pereira}, the GLLP type security proofs involving the quantum coin idea \citep{gottesman,lo4,koashi2} and the numerical techniques recently introduced in \cite{wang3,coles,winick}.

Furthermore, we have proposed a new framework for security proofs, which we call the reference technique (RT). It uses reference states that are similar to the states sent in the actual protocol, thus allowing us to determine the parameters needed to prove the security of the latter. The RT is very general and it can be applied to many QKD protocols. Moreover, it already includes the LT protocol, the GLT protocol and the GLLP type security proofs as special cases. That is, we are able to re-construct these security proofs by applying the RT, as shown in the Supplementary Material. Importantly, we have demonstrated that most of the source imperfections can be incorporated simultaneously into the RT and therefore, this technique has been proven to be very useful for guaranteeing the security of practical QKD protocols. In particular, we have shown that for the RT based on the original LT no information about the side-channel states is required, yet it is an analytical security proof, resulting in a much simpler characterisation of the source. Also, we emphasise that the RT can be applied together with analytical or numerical optimisation to estimate an upper bound on the phase error rate, which could result in a higher performance. In this work, we have rigorously proven the security of the RT and we have provided the sufficient conditions to apply this technique to other QKD protocols (see the Supplementary Material). We remark that for the security proof we have not considered the probabilities to be conditional on the detection events, which is usually important for high performance in the finite-key scenario. Fortunately, thanks to the recently developed Kato's inequality \cite{kato}, this is not a problem anymore, and it does not affect the performance of the secret key rate even in the finite-key size regime.

Additionally, in the Supplementary Material, we have compared the RT based on the original LT protocol with the RT based on the GLLP type security proofs. We remark, however, that this comparison might be considered unfair because the RT based on the GLLP type security proofs requires four states and, analytical or numerical optimisation. Finally, we note that if a better inequality to evaluate the deviation between the probabilities associated with the reference states and those associated with the actual states is available, then it could replace the inequality in  Eq.~(\ref{eq:bound1}), resulting in even higher secret key rates for the RT. Also, our method could be applied to other problems in quantum information theory where one needs to estimate summed probabilities. In this sense, our work not only proves the security of practical QKD systems, but also has a potential to contribute to quantum information theory in general.

\section{Materials and Methods}
\label{sec:methods}
\subsection{Reference technique}
\label{sec:new proof}

The RT is a new framework to prove the security of QKD protocols. It is general and can reproduce the GLLP type security proofs involving the quantum coin idea \cite{gottesman,lo4,koashi2} and the original LT protocol \cite{tamaki}. Moreover, it can be applied to many different protocols. To see this, we refer the reader to the Supplementary Material where we demonstrate that the GLLP type security proofs can be reconstructed from the RT. Additionally, we outline the sufficient conditions to use the RT and prove the security of an $m$-state protocol. In this section, however, we outline the key idea of the RT and show that it can be seen as a generalisation of the LT protocol. For concreteness of the explanation, we concentrate on a particular example, the three-state protocol with side channels.

Usually, to prove the security of QKD protocols, a relationship among the probabilities associated with the actual states needs to be established. Quite often, it is not straightforward to construct such {a} relationship, and the RT could be very useful to overcome this difficulty. The key idea is to consider a set of states, which we call the reference states, instead of the actual states. These reference states can be chosen freely, but they should be selected such that it is easy to derive a relationship among the {probabilities associated with them.} For this, it may be convenient to select the reference states in a structured space, such as a qubit space, and importantly, it is preferential that the resulting relationship is resilient against some imperfections in the space, such as the SPFs. Note that, this relationship is associated with the reference states and therefore it cannot be used directly in the security proof. However, since the reference states are chosen to be similar to the actual states, we can obtain a relationship associated with the actual states by slightly modifying the relationship for the reference states. In summary, the RT consists mainly of two parts: 
\begin{enumerate}
\item Reference formula part: Here we construct a relationship among the probabilities associated with the reference states;
\item Deviation evaluation part: Here we transform the relationship for the reference states into a relationship for the actual states by evaluating the deviation between the probabilities associated with the reference states and  those associated with the actual states. 
\end{enumerate}
We emphasise that, the reference states are purely a mathematical tool to construct the Reference formula, and we do not need to consider or imagine their practical implementation. Below, we show how to apply the RT in practice by presenting a rigorous security proof against coherent attacks for the three-state protocol. \\

\noindent {\textit{Security proof of the three-state protocol with side channels}} \\
Let us assume a three-state protocol where Alice chooses a normalised state $\ket{\psi_{j}}_{B}$ from the set $\{\ket{\psi_{j}}_{B}\}_{j=0_Z, 1_Z, 0_X}$ with probability $p_{j}$, for each pulse emission. For simplicity of discussion, we {assume} $p_{0_Z}=p_{1_Z}$. The assumptions on Bob's side have been described in the Results section. Namely, he measures the incoming pulses in the $Z$ or in the $X$ basis with probabilities $p_{Z_B}$ and $p_{X_B}$, respectively. More precisely, Bob's $Z$-basis ($X$-basis) measurement is represented by the POVM ${\{{\hat m}_{0_Z}}, {\hat m}_{1_Z}, {\hat m}_{f}\}$ $(\{{\hat m}_{0_X}, {\hat m}_{1_X}, {\hat m}_{f}\})$ and it satisfies the basis independent detection efficiency condition. Note that, in this protocol the key is generated from a subset of the states indexed by $j=0_Z, 1_Z$, i.e.~the $Z$ basis {and the bit values obtained by Bob's $Z$-basis measurement.} 

Now, we write the states sent by Alice in the form of Eq.~(\ref{eq:main}). That is, we expand the states $\ket{\psi_{j}}_{B}$ by using an orthonormal basis, and in doing so we select a qubit space that is common over the three states. This suggests that $\ket{\psi_{j}}_{B}$ can be, most generally, decomposed into
\begin{eqnarray}
\ket{\psi_{0_Z}}_{B}&=&(1-\epsilon_{0_Z})\ket{\phi_{0_Z}}_{B}+ \sqrt{1-(1-\epsilon_{0_Z})^2}\ket{\phi_{0_Z}^{\perp}}_{B},\nonumber\\
\ket{\psi_{1_Z}}_{B}&=&(1-\epsilon_{1_Z})\ket{\phi_{1_Z}}_{B}+ \sqrt{1-(1-\epsilon_{1_Z})^2}\ket{\phi_{1_Z}^{\perp}}_{B},\nonumber\\
\ket{\psi_{0_X}}_{B}&=&(1-\epsilon_{0_X})\ket{\phi_{0_X}}_{B}+ \sqrt{1-(1-\epsilon_{0_X})^2}\ket{\phi_{0_X}^{\perp}}_{B}\,,
\label{example protocol-method}
\end{eqnarray} 
where the state $\ket{\phi_j}_{B}$ represents the qubit part of the state $\ket{\psi_{j}}_{B}$ and the state $\ket{\phi_{j}^{\perp}}_{B}$ is a (possibly) unknown side-channel state that lives in any Hilbert space and it is orthogonal to $\ket{\phi_{j}}_{B}$. We stress that, this orthogonality is needed only for each setting choice $j$, but not between different choices of $j$. Examples of these states were presented in Eq.~(\ref{eq:act_bef_new}), however for generality we do not restrict ourselves only to that scenario. Note that if the side channel states include pulse correlations, with maximum correlation length $i$, then we need to consider $(i+1)$ virtual protocols and apply the RT to each one of them individually in order to guarantee their security, and consequently, the security of the three-state protocol (see the Results section for more details). In that case, the following discussion holds for any of these virtual protocols. 

In the security proof, we assume that the qubit parts $\{\ket{\phi_j}_B\}_{j=0_Z,1_Z,0_X}$, which are to be adopted as the reference states, are perfectly characterised and stable in time, but we do not require any knowledge about the side-channel states $\{\ket{\phi_{j}^{\perp}}_B\}_{j=0_Z,1_Z,0_X}$. From an experimental viewpoint, the unnecessity of characterising the side-channel state $\ket{\phi_j^\perp}_B$ in Eq.~(\ref{example protocol-method}) is a great advantage as in practice it is very challenging to perform measurements on arbitrary physical degrees of freedom. In {Eq.~(\ref{example protocol-method}),} the coefficient $\epsilon_{j}$ satisfying $0\le\epsilon_j \le1$ quantifies the deviation of the state $\ket{\psi_{j}}_{B}$ ($j\in\{0_Z, 1_Z, 0_X\}$) from the qubit space. That is, the states $\ket{\psi_{j}}_{B}$ are ideally qubit states, however, due to the presence of side channels, such as THA or pulse correlations, they deviate from this perfect scenario. Note that, if $\epsilon_{j}=1$ it means that it is impossible to find such a qubit space for  particular $j$. {In the security proof, we assume that we know an upper bound $\epsilon$ on $\epsilon_j$, i.e.~$\epsilon_j \le \epsilon$ for all $j$. Furthermore, we shall assume that the qubit states $\ket{\phi_j}_B$ are those defined in Eq.~(\ref{eq:alice_states_new}). To summarise,} even if there is no information about the side-channel state $\ket{\phi_{j}^{\perp}}_B$, our security proof works as long as we adopt $\ket{\phi_j}_B$ as the reference states and they are perfectly characterised and stable in time, and we know $\epsilon$ (or more generally, $\epsilon_j$). In particular, this means that the state $\ket{\phi_{j}^{\perp}}_B$ can vary in time and {can be dependent on the previous pulses,} and therefore, the states $\ket{\psi_j}_B$ emitted by Alice's source do not need to be regarded as independently and identically distributed. This point will become clearer after Eq.~(\ref{eq:refeq}). We remark, however, that if we select the reference states containing side-channel states they will no longer be perfectly known or stable in time. In this case, to make the mathematical analysis simpler one could use analytical or numerical optimisation to consider the worst case scenario for the side-channel states, i.e.~the case that maximises the phase error rate. This maximisation removes the potential dependence on the previous pulses and thus effectively provides pulses that are independent and stable in time. Importantly, this is a purely  mathematical step and it does not require any extra assumptions on Alice's source, e.g., the states $\ket{\psi_j}_B$ do not need to be regarded as independently and identically distributed. 

Having finished the description of the states, we move on to the security proof using the RT. We are interested in proving the  security of the bit values generated from the $Z$-basis events. From Eve's perspective, this instance is equivalent to the one in which Alice selects the $Z$ basis, prepares systems $A$ and $B$ in the state
\begin{eqnarray}
\frac{1}{\sqrt{2}}\big(\ket{0_Z}_{A}\ket{\psi_{0_Z}}_{B}+ \ket{1_Z}_{A}\ket{\psi_{1_Z}}_{B}\big),
\label{eq:joint}
\end{eqnarray} 
sends system $B$ to Bob while keeping system $A$ in her lab, and then both Alice and Bob perform their measurements in the $Z$ basis. To prove the security of the $Z$-basis events, we need to estimate the phase errors \cite{shor,tamaki}, which are defined in the $X$ basis. That is, we consider the errors that Alice and Bob would have obtained if Alice had performed  the $X$-basis measurement $\{\ket{0_X}_{A}, \ket{1_X}_{A}\}$ (with $\ket{0_X}_{A}:=({\ket{0_Z}_{A}+\ket{1_Z}_{A}})/{\sqrt{2}}$ and $\ket{1_X}_{A}:=({\ket{0_Z}_{A}-\ket{1_Z}_{A}})/{\sqrt{2}}$) and Bob had employed {a basis complementary to the $Z$ basis} (a suitable choice under the basis independent efficiency condition, may be the $X$ basis used in the actual protocol) for the measurement on the joint state defined in Eq.~(\ref{eq:joint}). This leads us to consider a virtual protocol in which Alice sends the virtual states $\ket{\psi_{0_X}^{\rm vir}}_{B}\propto \ket{\psi_{0_Z}}_{B}+\ket{\psi_{1_Z}}_{B}$ and $\ket{\psi_{1_X}^{\rm vir}}_{B}\propto \ket{\psi_{0_Z}}_{B}-\ket{\psi_{1_Z}}_{B}$ \cite{tamaki} to Bob with probabilities
\begin{align}
{\tilde p}_{\alpha_X}^{\rm vir} =  \frac{1}{2} {p}_{Z_A} \big[{1+(-1)^\alpha \Re( {}_{B}\bra{\psi_{0_Z}}\psi_{1_Z}\rangle_{B}})\big],
\end{align}where $p_{Z_A}$ is the probability that Alice selects the $Z$ basis. Here, ${\tilde p}_{0_X}^{\rm vir}$ (${\tilde p}_{1_X}^{\rm vir}$) is the joint probability that Alice selects the $Z$ basis and prepares the normalised virtual state $\ket{\psi_{0_X}^{\rm vir}}_{B}$ ($\ket{\psi_{1_X}^{\rm vir}}_{B}$), through the $X$-basis measurement.

In the security proof, it is convenient to represent the actual protocol in terms of a virtual entanglement-based protocol. As explained above, in this virtual protocol we consider replacing Alice and Bob's bases with the $X$ basis when both of them select the $Z$ basis. From Eve's viewpoint, the actual protocol with this replacement can be equivalently described by Alice and Bob  fictitiously preparing the following entangled state 
\begin{eqnarray}
\ket{\psi}_{CB}&:=&\sqrt{p_{0_Z}p_{X_B}}\ket{0_{z,A}, X_{B}}_{C}\ket{\psi_{0_Z}}_{B}+\sqrt{p_{1_Z}p_{X_B}}\ket{1_{z,A}, X_{B}}_{C}\ket{\psi_{1_Z}}_{B}\nonumber\\
&+&\sqrt{p_{0_X}p_{X_B}}\ket{0_{x,A}, X_{B}}_{C}\ket{\psi_{0_X}}_{B}+\sqrt{p_{0_X}p_{Z_B}}\ket{0_{x,A}, Z_{B}}_{C}\ket{\psi_{0_X}}_{B}\nonumber\\
&+&\sqrt{{\tilde p}_{0_X}^{\rm vir}p_{Z_B}}\ket{0_{x,A}^{\rm vir}, X_{B}}_{C}\ket{\psi_{0_X}^{\rm vir}}_{B}+\sqrt{{\tilde p}_{1_X}^{\rm vir}p_{Z_B}}\ket{1_{x,A}^{\rm vir}, X_{B}}_{C}\ket{\psi_{1_X}^{\rm vir}}_{B}\,,
\label{example-actual-entanglement-based-protocol-method}
\end{eqnarray}
and then performing a measurement on system $C$, which is associated with Alice's and Bob's {setting} choices. In particular, $\ket{0_{z,A}, X_{B}}_{C}$ ($\ket{1_{x,A}^{\rm vir}, X_{B}}_{C}$) represents the events when Alice selects the actual state for $0_Z$ (the virtual state for $1_X$) and Bob chooses the $X$ basis. Note that, there are six states of system $C$ which store different classical information related with Alice's and Bob's setting choices, and they are all normalised and orthogonal to each other. After Alice prepares the entangled state in Eq.~(\ref{example-actual-entanglement-based-protocol-method}), we imagine that she performs an orthogonal measurement that projects system $C$ onto one of these six states, {and Bob performs the measurement according to the basis directed by the measurement outcome.} We remark that, the first four terms in Eq.~(\ref{example-actual-entanglement-based-protocol-method}) correspond to the actual events that occur in the actual protocol, while the last two terms correspond to the virtual events. That is, the last two terms represent the events in which Alice and Bob select the $Z$ basis in the actual protocol, however, their basis choice is replaced by the $X$ basis for the security proof. Importantly, the virtual events and the actual events are clearly defined in system $C$, and they correspond to disjoint events. The actual protocol can then be regarded as repeatedly, say $N$ times, preparing systems $B$ and $C$ in the state $\ket{\psi}_{CB}$ followed by the measurements by Alice and Bob. Now, following the steps of the RT introduced above we can estimate the phase errors associated with this protocol.

\begin{enumerate}
\item \underline{Reference formula part}

As an example, we choose the reference states to be the qubit part of the actual states. For the actual states defined in Eq.~(\ref{example protocol-method}), this corresponds to selecting the set $\{\ket{\phi_{0_Z}}_{B}, \ket{\phi_{1_Z}}_{B}, \ket{\phi_{0_X}}_{B}\}$ (see Eq.~(\ref{eq:alice_states_new}) for their explicit form). Now, we need to construct a relationship associated with these reference states. First, we consider the virtual entangled state
\begin{eqnarray}
\frac{1}{\sqrt{2}}\big(\ket{0_Z}_{A}\ket{\phi_{0_Z}}_{B}+ \ket{1_Z}_{A}\ket{\phi_{1_Z}}_{B}\big).
\label{eq:joint_ref}
\end{eqnarray} 
Note that, Eq.~(\ref{eq:joint_ref}) is analogous to Eq.~(\ref{eq:joint}) but the actual states have been replaced with their respective reference states. Then, we may imagine that Alice measures system $A$ in the $X$ basis and sends Bob the virtual states $\ket{\phi_{0_X}^{\rm vir}}_{B}\propto \ket{\phi_{0_Z}}_{B}+\ket{\phi_{1_Z}}_{B}$ and $\ket{\phi_{1_X}^{\rm vir}}_{B}\propto \ket{\phi_{0_Z}}_{B}-\ket{\phi_{1_Z}}_{B}$ with probabilities
\begin{eqnarray}
p_{\alpha_X}^{\rm vir} = \frac{1}{2} {p}_{Z_A} \big[{1+(-1)^\alpha \Re ({}_{B}\bra{\phi_{0_Z}}\phi_{1_Z}\rangle_{B}})\big].
\label{eq:refprob}
\end{eqnarray} Here, $p_{0_X}^{\rm vir}$ ($p _{1_X}^{\rm vir}$) could be interpreted as the joint probability that Alice selects the $Z$ basis and prepares the normalised virtual state $\ket{\phi_{0_X}^{\rm vir}}_{B}$ ($\ket{\phi_{1_X}^{\rm vir}}_{B}$), through the $X$-basis measurement. Now, we mathematically replace all the actual and virtual states in Eq.~(\ref{example-actual-entanglement-based-protocol-method}) with their respective reference states:
\begin{eqnarray}
&&\sqrt{p_{0_Z}p_{X_B}}\ket{0_{z,A}, X_{B}}_{C}\ket{\phi_{0_Z}}_{B}+\sqrt{p_{1_Z}p_{X_B}}\ket{1_{z,A}, X_{B}}_{C}\ket{\phi_{1_Z}}_{B}\nonumber\\
&+&\sqrt{p_{0_X}p_{X_B}}\ket{0_{x,A}, X_{B}}_{C}\ket{\phi_{0_X}}_{B}+\sqrt{p_{0_X}p_{Z_B}}\ket{0_{x,A}, Z_{B}}_{C}\ket{\phi_{0_X}}_{B}\nonumber\\
&+&\sqrt{p_{0_X}^{\rm vir}p_{Z_B}}\ket{0_{x,A}^{\rm vir}, X_{B}}_{C}\ket{\phi_{0_X}^{\rm vir}}_{B}+\sqrt{p_{1_X}^{\rm vir}p_{Z_B}}\ket{1_{x,A}^{\rm vir}, X_{B}}_{C}\ket{\phi_{1_X}^{\rm vir}}_{B}.
\label{example-reference-entanglement-based-protocol-method}
\end{eqnarray}
Again, we emphasise that this entanglement-based protocol with the reference states is purely a mathematical tool for the security proof, and we do {\it not} need to consider or imagine its practical implementation.

The reason why we have selected $\{\ket{\phi_{0_Z}}_{B}, \ket{\phi_{1_Z}}_{B}, \ket{\phi_{0_X}}_{B}\}$ as the reference states is two-fold. First, these states are close to their respective actual states $\{\ket{\psi_{0_Z}}_{B}, \ket{\psi_{1_Z}}_{B}, \ket{\psi_{0_X}}_{B}\}$. Therefore, we expect that the probabilities associated with the reference states should be similar to those associated with the actual states. Second, by directly employing the idea of the LT protocol for a qubit-based protocol \cite{tamaki}, we can obtain a relationship between the reference states and the virtual states, which is expected to be {loss-tolerant against SPFs.} More concretely, below we consider that the reference states used are the ones defined in Eq.~(\ref{eq:alice_states_new}), and, in this case, we can express the virtual states for the reference states as in Eq.~(\ref{eq:qspace}). We rewrite it here for convenience,
\begin{eqnarray}
&& \dyad{\phi_{1_X}^{\rm vir}}{\phi_{1_X}^{\rm vir}}_B = a \dyad{\phi_{0_Z}}{\phi_{0_Z}}_B + b \dyad{\phi_{1_Z}}{\phi_{1_Z}}_B - c \dyad{\phi_{0_X}}{\phi_{0_X}}_B,\nonumber \\
&& \dyad{\phi_{0_X}^{\rm vir}}{\phi_{0_X}^{\rm vir}}_B =  \dyad{\phi_{0_X}}{\phi_{0_X}}_B, \label{exploit-qubit-method}
\end{eqnarray}
{where from Eq.~(\ref{eq:alice_states_new}), the coefficients $a, b,c \ge 0$ are given by}
\begin{align}
&a := \frac{-2\sin(\frac{\pi}{4}+ \frac{\delta}{4})}{\cos(\frac{\pi}{4} + \frac{3\delta}{4}) - 3\sin(\frac{\pi}{4} + \frac{\delta}{4})}, \nonumber \\
&b :=  \frac{-2\sin(\frac{\pi}{4}+ \frac{\delta}{4})}{\cos(\frac{\pi}{4} + \frac{3\delta}{4}) - 3\sin(\frac{\pi}{4} + \frac{\delta}{4})}, \nonumber \\
&c: = \frac{-\sin(\frac{\delta}{2}) + 1}{\sin(\frac{\delta}{2}) + 1}.
\end{align} We remark that, in Eq.~(\ref{exploit-qubit-method}), we have highly exploited the properties of a qubit space, i.e.~even with a negative sign in front of the coefficient $c$,  $\dyad*{\phi_{1_X}^{\rm vir}}{\phi_{1_X}^{\rm vir}}_{B}$ is still a density operator, which would not be the case in general for a density operator in a Hilbert space with a higher dimension.

We now consider the following quantity,
\begin{eqnarray}
P^{(k)}({\rm ph|Ref}):=p_{1_X}^{\rm vir} p_{Z_B}{\rm Tr}\big[\dyad*{\phi_{1_X}^{\rm vir}}{\phi_{1_X}^{\rm vir}}_B {\hat M}_{0_X}^{(k)}\big]+p_{0_X}^{\rm vir} p_{Z_B}{\rm Tr}\big[\dyad*{\phi_{0_X}^{\rm vir }}{\phi_{0_X}^{\rm vir}}_B{\hat M}_{1_X}^{(k)}\big]\,,
\label{phase-error-reference-method}
\end{eqnarray}
which, as described in the Results section, could be interpreted as the probability of a phase error for the $k^{\rm th}$ pulse when employing the reference states, and where ${\hat M}_{\alpha_X}^{(k)}$ with $\alpha\in \{0,1\}$ is Bob's POVM element for the $k^{\rm th}$ pulse after a coherent attack in the {\it actual} protocol, that is {${\hat M}_{\alpha X}^{(k)}:= \sum_{\tilde e} {\hat K}_{\tilde e}^{(k)} {\hat m}_{\alpha_X}{\hat K}_{\tilde e}^{(k)\dagger}$.} Here, {${\hat K}_{\tilde e}^{(k)}$} is the Kraus operator representing the action that the $k^{\rm th}$ pulse is subjected to. This operator is obtained by Eve's coherent attack that acts on the $k^{\rm th}$ pulse sent by Alice  by considering all the $k-1$ previous measurements by Alice and Bob. {Here, ${\tilde e}$ represents a particular outcome of the measurement conducted by Eve.} The goal now is to transform the quantities associated with the reference states in Eq.~(\ref{phase-error-reference-method}) into those associated with the actual states for the $k^{\rm th}$ pulse:
\begin{eqnarray}
P^{(k)}({\rm ph|Act}):={\tilde p}_{1_X}^{\rm vir} p_{Z_B}{\rm Tr}\big[\dyad*{\psi_{1_X}^{\rm vir}}{\psi_{1_X}^{\rm vir}}_B{\hat M}_{0_X}^{(k)}\big]+{\tilde p}_{0_X}^{\rm vir} p_{Z_B}{\rm Tr}\big[\dyad*{\psi_{0_X}^{\rm vir}}{\psi_{0_X}^{\rm vir}}_B{\hat M}_{1_X}^{(k)}\big]\,.
\label{phase-error-actual-method}
\end{eqnarray}
Using Eq.~(\ref{exploit-qubit-method}), we can express Eq.~(\ref{phase-error-reference-method}) as \newpage
\begin{align}
P^{(k)}({\rm ph|Ref})&=p_{1_X}^{\rm vir} p_{Z_B}a{\rm Tr}\big[\dyad*{\phi_{0_Z}}{\phi_{0_Z}}_B{\hat M}_{0_X}^{(k)}\big]+p_{1_X}^{\rm vir} p_{Z_B}b{\rm Tr}\big[\dyad*{\phi_{1_Z}}{\phi_{1_Z}}_B{\hat M}_{0_X}^{(k)}\big]\nonumber\\
&-p_{1_X}^{\rm vir} p_{Z_B}c{\rm Tr}\big[\dyad*{\phi_{0_X}}{\phi_{0_X}}_B{\hat M}_{0_X}^{(k)}\big]+p_{0_X}^{\rm vir} p_{Z_B}{\rm Tr}\big[\dyad*{\phi_{0_X}}{\phi_{0_X}}_B{\hat M}_{1_X}^{(k)}\big]\,,
\end{align}
which is equivalent to 
\begin{align}
0&=p_{1_X}^{\rm vir} p_{Z_B}a{\rm Tr}\big[\dyad*{\phi_{0_Z}}{\phi_{0_Z}}_B{\hat M}_{0_X}^{(k)}\big]+p_{1_X}^{\rm vir} p_{Z_B}b{\rm Tr}\big[\dyad*{\phi_{1_Z}}{\phi_{1_Z}}{\hat M}_{0_X}^{(k)}\big]+p_{0_X}^{\rm vir} p_{Z_B}{\rm Tr}\big[\dyad*{\phi_{0_X}}{\phi_{0_X}}_B{\hat M}_{1_X}^{(k)}\big]\nonumber\\
&-\Big(p_{1_X}^{\rm vir} p_{Z_B}c{\rm Tr}\big[\dyad*{\phi_{0_X}}{\phi_{0_X}}_B{\hat M}_{0_X}^{(k)}\big]+p_{1_X}^{\rm vir} p_{Z_B}{\rm Tr}\big[\dyad*{\phi_{1_X}^{\rm vir}}{\phi_{1_X}^{\rm vir}}_B{\hat M}_{0_X}^{(k)}\big]+p_{0_X}^{\rm vir} p_{Z_B}{\rm Tr}\big[\dyad{\phi_{0_X}^{\rm vir}}{\phi_{0_X}^{\rm vir}}{\hat M}_{1_X}^{(k)}\big]\Big)\,,
\label{example-phase-error-formula}
\end{align}
Here, we emphasise that Eq.~(\ref{example-phase-error-formula}) is derived based on the idea of the LT protocol, and therefore it entails the robustness against the SPFs in the qubit space. That is, if there are no side channels, i.e.~$\epsilon=0$, then Eq.~(\ref{example-phase-error-formula}), which is exactly the expression that is used in the original LT protocol \cite{tamaki}, results in a secret key rate that is loss-tolerant against SPFs. Therefore, this shows that the RT includes the LT protocol in the Reference formula part. Next, we transform the relationship for the reference states in Eq.~(\ref{example-phase-error-formula}) into a relationship for the actual states. That is, we enter the Deviation evaluation part of the RT.

\item \underline{Deviation evaluation part}

For the transformation of Eq.~(\ref{example-phase-error-formula}), we employ the bound in Eq.~(\ref{eq:bound1}). We rewrite it here for convenience: 
{\begin{eqnarray}
g^L \Big( {\rm Tr}\big[\dyad{A}{A}{\hat M}\big],|\bra{A}{R}\rangle| \Big) \le {\rm Tr}\big[\dyad{R}{R}{\hat M}\big] \le g^U \Big( {\rm Tr}\big[\dyad{A}{A}{\hat M}\big],|\bra{A}{R}\rangle| \Big),
\label{main-bound-method}
\end{eqnarray}}where $\ket{R}$ ($\ket{A}$) is any normalised state associated with the reference (actual) states and 
{\begin{align}
&g^L(x,y) = \left\{
        \begin{array}{ll}
            0 & \quad x  < 1 - y^2  \\
            x + (1-y^2)(1-2x) - 2y\sqrt{(1-y^2)x(1-x)} & \quad x  \geq 1 - y^2, 
        \end{array}
    \right. \\
&g^U(x,y) = \left\{
        \begin{array}{ll}
            x + (1-y^2)(1-2x) + 2y\sqrt{(1-y^2)x(1-x)} & \quad x \leq y^2  \\
            1 & \quad x > y^2.
        \end{array}
    \right.
\end{align}}Note that, {$-g^{L}(x,y)$ and $g^{U}(x,y)$} are concave with respect to $0\le x\le1$ for any fixed $0\le y\le1$. For more details on the derivation of Eq.~(\ref{main-bound-method}) see the Supplementary Material. Now, we consider the first three terms in Eq.~(\ref{example-phase-error-formula}), which are re-expressed as
\begin{eqnarray}
p_{Z_B} S_{+} {\rm Tr}\left[\frac{p_{1_X}^{\rm vir}a}{S_{+}} \dyad*{\phi_{0_Z}}{\phi_{0_Z}}_B {\hat M}_{0_X}^{(k)}+ \frac{p_{1_X}^{\rm vir}b}{S_{+}}\dyad*{\phi_{1_Z}}{\phi_{1_Z}}_B{\hat M}_{0_X}^{(k)}+\frac{p_{0_X}^{\rm vir}}{S_{+}}\dyad*{\phi_{0_X}}{\phi_{0_X}}_B{\hat M}_{1_X}^{(k)}\right],
\label{eq:1stPart}
\end{eqnarray}
where $S_{+}:=p_{1_X}^{\rm vir}a+p_{1_X}^{\rm vir}b+p_{0_X}^{\rm vir}$ is a normalisation factor. Next, we rewrite the term $\Tr[\cdot]$ in Eq.~(\ref{eq:1stPart}) as
\begin{align}
&{\rm Tr}\left[{\hat P}\left(\sqrt{\frac{p_{1_X}^{\rm vir}a}{S_{+}}}\ket{0_{z,A}, X_{B}}_{C}\ket{\phi_{0_Z}}_{B}+\sqrt{\frac{p_{1_X}^{\rm vir}b}{S_{+}}}\ket{1_{z,A}, X_{B}}_{C}\ket{\phi_{1_Z}}_{B}+\sqrt{\frac{p_{0_X}^{\rm vir}}{S_{+}}}\ket{0_{x,A}, X_{B}}_{C}\ket{\phi_{0_X}}_{B}\right){\hat M_{+}}^{(k)}\right]\nonumber\\
&=:{\rm Tr}\big[{\hat P}(\ket{R_{+}}_{CB}){\hat M_{+}}^{(k)}\big],
\label{M2-operator}
\end{align}
with 
\begin{align}
&\ket{R_{+}}_{CB}:=\sqrt{\frac{p_{1_X}^{\rm vir}a}{S_{+}}}\ket{0_{z,A}, X_{B}}_{C}\ket{\phi_{0_Z}}_{B}+\sqrt{\frac{p_{1_X}^{\rm vir}b}{S_{+}}}\ket{1_{z,A}, X_{B}}_{C}\ket{\phi_{1_Z}}_{B}+\sqrt{\frac{p_{0_X}^{\rm vir}}{S_{+}}}\ket{0_{x,A}, X_{B}}_{C}\ket{\phi_{0_X}}_{B},\nonumber\\
&{\hat M}^{(k)}_{+}:={\hat P}\left(\ket{0_{z,A}, X_{B}}_{C}\right)\otimes{\hat M}_{0_X}^{(k)}+{\hat P}\left(\ket{1_{z,A}, X_{B}}_{C}\right)\otimes{\hat M}_{0_X}^{(k)}+{\hat P}\left(\ket{0_{x,A}, X_{B}}_{C}\right)\otimes{\hat M}_{1_X}^{(k)}, \nonumber
\end{align}
where ${\hat P}(\ket{\cdot})=\dyad{\cdot}{\cdot}$.
Note that, this is purely a mathematical reinterpretation of the summed probabilities. We are interested in mathematically replacing $\ket{\phi_{0_Z}}_{B}$, $\ket{\phi_{1_Z}}_{B}$, and $\ket{\phi_{0_X}}_{B}$ in Eq.~(\ref{M2-operator}) with $\ket{\psi_{0_Z}}_{B}$, $\ket{\psi_{1_Z}}_{B}$, and $\ket{\psi_{0_X}}_{B}$, respectively, by employing Eq.~(\ref{main-bound-method}). For this, we may select
\begin{eqnarray}
\ket{A_{+}}_{CB}:=\sqrt{\frac{p_{1_X}^{\rm vir}a}{S_{+}}}\ket{0_{z,A}, X_{B}}_{C}\ket{\psi_{0_Z}}_{B}+\sqrt{\frac{p_{1_X}^{\rm vir}b}{S_{+}}}\ket{1_{z,A}, X_{B}}_{C}\ket{\psi_{1_Z}}_{B}+\sqrt{\frac{p_{0_X}^{\rm vir}}{S_{+}}}\ket{0_{x,A}, X_{B}}_{C}\ket{\psi_{0_X}}_{B},
\end{eqnarray}
and as a result, we have transformed the first three terms of Eq.~(\ref{example-phase-error-formula}) into
\begin{align}
&p_{1_X}^{\rm vir} p_{Z_B}a{\rm Tr}\big[\dyad*{\phi_{0_Z}}{\phi_{0_Z}}_B{\hat M}_{0_X}^{(k)}\big]+p_{1_X}^{\rm vir} p_{Z_B}b{\rm Tr}\big[\dyad*{\phi_{1_Z}}{\phi_{1_Z}}{\hat M}_{0_X}^{(k)}\big]+p_{0_X}^{\rm vir} p_{Z_B}{\rm Tr}\big[\dyad*{\phi_{0_X}}{\phi_{0_X}}_B{\hat M}_{1_X}^{(k)}\big]\nonumber\\
& {\le} p_{Z_B} S_{+} {g^U} \left({\rm Tr}\big[{\hat P}(\ket{A_{+}}_{CB}){\hat M}^{(k)}_{+}\big],{1-\epsilon}\right),
\label{result^1st}
\end{align}
{where we have selected an upper bound on $\Tr \big [\hat{P}(\ket{R_{+}}_{CB}) \hat{M}_{+}^{(k)} \big]$ to obtain an upper bound on the phase error probability, and} used $|{}_{CB}\langle A_{+}\ket{R_{+}}_{CB}| = 1-\epsilon$. Here, note that in order to calculate this inner product we need to calculate the terms ${}_{B}\langle \psi_{j}\ket{\phi_{j}}_{B}$ {rather than} ${}_{B}\langle \psi_{j}\ket{\phi_{\tilde j}}{}_{B}$ with $j \neq \tilde j$, {which shows the aforementioned simplicity of the state characterisation needed in our proof. Now,} to clearly see how the term ${\rm Tr}\big[{\hat P}(\ket{A_{+}}_{CB}){\hat M}^{(k)}_{+}\big]$ is related with the quantities obtained from an experimental implementation of the actual protocol, we write
\begin{align}
&{\rm Tr}\big[{\hat P}(\ket{A_{+}}_{CB}){\hat M}^{(k)}_{+}\big]\nonumber\\
&= {\rm Tr}\left[\frac{p_{1_X}^{\rm vir}a}{S_{+}} \dyad*{\psi_{0_Z}}{\psi_{0_Z}}{\hat M}_{0_X}^{(k)}+ \frac{p_{1_X}^{\rm vir}b}{S_{+}}\dyad*{\psi_{1_Z}}{\psi_{1_Z}}{\hat M}_{0_X}^{(k)}+\frac{p_{0_X}^{\rm vir}}{S_{+}}\dyad*{\psi_{0_X}}{\psi_{0_X}}{\hat M}_{1_X}^{(k)}\right] \nonumber \\
&=\frac{p_{1_X}^{\rm vir}a}{S_{+} {p_{0z}p_{X_B}}}{P^{(k)}(q_{0z,0x}|{\rm Act})}+\frac{p_{1_X}^{\rm vir}b}{S_{+} {p_{1z}p_{X_B}}}{P^{(k)}(q_{1z,0x}|{\rm Act})}+~\frac{p_{0_X}^{\rm vir}}{S_{+} {p_{0x}p_{X_B}}}{P^{(k)}(q_{0x,1x}|{\rm Act}).}
\label{eq:part2}
\end{align}
Here, ${P^{(k)}(q_{0z,0x}|{\rm Act})}$ is the joint probability that Alice selects the setting $0_Z$ and Bob's measurement outcome is $0_X$ at the $k^{\rm th}$ instance, conditional on the first $k-1$ measurements by Alice and Bob in the entanglement-based protocol for the actual protocol. The other probabilities are defined in a similar manner. This finishes the transformation of the first three terms with respect to the probabilities associated with the actual protocol.

Next, we consider the last three terms in Eq.~(\ref{example-phase-error-formula}), which are re-expressed as
\begin{eqnarray}
p_{Z_B} S_{-} {\rm Tr}\left[\frac{p_{1_X}^{\rm vir}c}{S_{-}} \dyad*{\phi_{0_X}}{\phi_{0_X}}_B{\hat M}_{0_X}^{(k)}+\frac{p_{1_X}^{\rm vir}}{S_{-}}\dyad*{\phi_{1_X}^{\rm vir}}{\phi_{1_X}^{\rm vir}}_B{\hat M}_{0_X}^{(k)}+\frac{p_{0_X}^{\rm vir}}{S_{-}}\dyad*{\phi_{0_X}^{\rm vir}}{\phi_{0_X}^{\rm vir}}_B{\hat M}_{1_X}^{(k)}\right],
\label{eq:pos_part}
\end{eqnarray}
where $S_{-}:=p_{1_X}^{\rm vir} c+p_{1_X}^{\rm vir}+p_{0_X}^{\rm vir}=p_{1_X}^{\rm vir} c + {p_{Z_A}}$, with ${p_{Z_A} := p_{0_Z} + p_{1_Z},}$ is the normalisation factor. The term $\Tr[\cdot]$ in  Eq.~(\ref{eq:pos_part}) can be expressed as
\begin{align}
&{\rm Tr}\left[{\hat P}\left(\sqrt{\frac{p_{1_X}^{\rm vir}c}{S_{-}}}\ket{0_{x,A}, X_{B}}_{C}\ket{\phi_{0_X}}_{B}+{\sqrt{\frac{p_{Z_A}}{2S_{-}}}\ket{0_{z,A}, Z_{B}}_{C}\ket{\phi_{0_Z}}_{B}}+{\sqrt{\frac{p_{Z_A}}{2 S_{-}}}\ket{1_{z,A}, Z_{B}}_{C}\ket{\phi_{1_Z}}_{B}}\right){\hat M}^{(k)}_{-}\right]\nonumber\\
&=:{\rm Tr}\big[{\hat P}(\ket{R_{-}}_{CB}){\hat M}^{(k)}_{-}\big],
\label{M-operator}
\end{align}
with
\begin{align}
&\ket{R_{-}}_{CB}:=\sqrt{\frac{p_{1_X}^{\rm vir}c}{S_{-}}}\ket{0_{x,A}, X_{B}}_{C}\ket{\phi_{0_X}}_{B}+{\sqrt{\frac{p_{Z_A}}{2S_{-}}}\ket{0_{z,A}, Z_{B}}_{C}\ket{\phi_{0_Z}}_{B}}+ {\sqrt{\frac{p_{Z_A}}{2 S_{-}}}\ket{1_{z,A}, Z_{B}}_{C}\ket{\phi_{1_Z}}_{B},} \nonumber\\
&{\hat M}^{(k)}_{-}:={\hat P}\left(\ket{0_{x,A}, X_{B}}_{C}\right)\otimes{\hat M}_{0_X}^{(k)}+{{\hat P}\left(\frac{\ket{0_{z,A}, Z_{B}}_{C} - \ket{1_{z,A}, Z_{B}}_{C}}{\sqrt{2}}\right)}\otimes{\hat M}_{0_X}^{(k)} \nonumber \\
&~~~~~~~~+{{\hat P}\left(\frac{\ket{0_{z,A}, Z_{B}}_{C} + \ket{1_{z,A}, Z_{B}}_{C}}{\sqrt{2}}\right)}\otimes{\hat M}_{1_X}^{(k)}.
\end{align}
Here, we have used the fact that the state $\ket{\phi_{\alpha_X}^{\rm vir}}_B = \frac{1}{2} (\ket{\phi_{0_Z}}_B + (-1)^\alpha \ket{\phi_{1_Z}}_B)/\sqrt{p_{\alpha_X}^{\rm vir}/p_{Z_A}}$ and that ${\hat P}\left(\ket{0_{z,A}, Z_{B}}_{C} - \ket{1_{z,A}, Z_{B}}_{C}/\sqrt{2}\right)$ and ${\hat P}\left(\ket{0_{z,A}, Z_{B}}_{C} + \ket{1_{z,A}, Z_{B}}_{C}/\sqrt{2}\right)$ correspond to the events associated with the states $\ket{1_{x,A}^{\rm vir}, X_B}$ and $\ket{0_{x,A}^{\rm vir}, X_B}$ in the entanglement-based virtual protocol, respectively. Again, we remark that this is purely a mathematical reinterpretation of the summed probability. In order to mathematically replace the states involving the reference states with those involving the actual states, we may select 
\begin{eqnarray}
\ket{A_{-}}_{CB}:=\sqrt{\frac{p_{1_X}^{\rm vir}c}{S_{-}}}\ket{0_{x,A}, X_{B}}_{C}\ket{\psi_{0_X}}_{B}+{\sqrt{\frac{{p_{Z_A}}}{2S_{-}}}\ket{0_{z,A}, Z_{B}}_{C}\ket{\psi_{0_Z}}_{B}}+{\sqrt{\frac{{ p_{Z_A}}}{2 S_{-}}}\ket{1_{z,A}, Z_{B}}_{C}\ket{\psi_{1_Z}}_{B}}.
\label{actual-A_{-}}
\end{eqnarray}
As a result, we have transformed the last three terms in Eq. (\ref{example-phase-error-formula}) into \newpage
\begin{align}
&p_{1_X}^{\rm vir} p_{Z_B}c{\rm Tr}\big[\dyad*{\phi_{0_X}}{\phi_{0_X}}_B{\hat M}_{0_X}^{(k)}\big]+p_{1_X}^{\rm vir} p_{Z_B}{\rm Tr}\big[\dyad*{\phi_{1_X}^{\rm vir}}{\phi_{1_X}^{\rm vir}}_B{\hat M}_{0_X}^{(k)}\big]+p_{0_X}^{\rm vir} p_{Z_B}{\rm Tr}\big[\dyad*{\phi_{0_X}^{\rm vir}}{\phi_{0_X}^{\rm vir}}_B{\hat M}_{1_X}^{(k)}\big]\nonumber\\
& {\ge} p_{Z_B} S_{-} {g^L} \left({\rm Tr}\big[{\hat P}(\ket{A_{-}}_{CB}){\hat M}^{(k)}_{-}\big],{1-\epsilon}\right),
\label{result^2nd}
\end{align}
{where we have selected a lower bound on $\Tr \big [\hat{P}(\ket{R_{-}}_{CB}) \hat{M}_{-}^{(k)} \big]$ to obtain an upper bound on the phase error probability, and used} $|{_{CB}}\langle A_{-}\ket{R_{-}}_{CB}| = 1-\epsilon$. As before, in order to calculate this inner product we need to calculate the terms ${}_{B}\langle \psi_{j}\ket{\phi_{j}}_{B}$ {rather than} ${}_{B}\langle \psi_{j}\ket{\phi_{\tilde j}}_{B}$ with $j \neq \tilde j$. Now, we look at ${\rm Tr}\big[{\hat P}(\ket{A_{-}}_{CB}){\hat M}^{(k)}_{-}\big]$, which is expressed and interpreted by
\begin{align}
&{\rm Tr}\Big[{\hat P}(\ket{A_{-}}_{CB}){\hat M}^{(k)}_{-}\Big]\nonumber\\
&{= \Tr \left[ \hat{P} \left( \sqrt{\frac{p_{1_X}^{\rm vir}c}{S_{-}}}\ket{0_{x,A}, X_{B}}_{C}\ket{\psi_{0_X}}_{B}+{\sqrt{\frac{{p_{Z_A}}}{2S_{-}}}\ket{0_{z,A}, Z_{B}}_{C}\ket{\psi_{0_Z}}_{B}}+{\sqrt{\frac{{ p_{Z_A}}}{2 S_{-}}}\ket{1_{z,A}, Z_{B}}_{C}\ket{\psi_{1_Z}}_{B}} \right) \hat M_{-}^{(k)} \right]}\nonumber \\
&={\rm Tr}\left[\frac{p_{1_X}^{\rm vir}c}{S_{-}} \dyad*{\psi_{0_X}}{\psi_{0_X}}_B{\hat M}_{0_X}^{(k)}+\frac{{\tilde p_{1_X}}^{\rm vir}}{S_{-}}\dyad*{\psi_{1_X}^{\rm vir}}{\psi_{1_X}^{\rm vir}}_B{\hat M}_{0_X}^{(k)}+\frac{{\tilde p_{0_X}}^{\rm vir}}{S_{-}}\dyad*{\psi_{0_X}^{\rm vir}}{\psi_{0_X}^{\rm vir}}_B{\hat M}_{1_X}^{(k)}\right] \nonumber \\
&=\frac{p_{1_X}^{\rm vir}c}{S_{-}{p_{0_X}p_{X_B}}}P^{(k)}(q_{0x,0x}|{\rm Act})+\frac{1}{S_{-}p_{Z_B}}P^{(k)}({\rm ph|Act})\,,
\label{eq:realprob}
\end{align}
where we have used Eq. (\ref{phase-error-actual-method}), namely, the definition of $P^{(k)}({\rm ph|Act})$ and the fact that the states $\ket{\psi_{\alpha_X}^{\rm vir}}_B = \frac{1}{2} (\ket{\psi_{0_Z}}_B + (-1)^\alpha \ket{\psi_{1_Z}}_B)/\sqrt{\tilde p_{\alpha_X}^{\rm vir}/p_{Z_A}}$.

Now, we combine Eqs.~(\ref{example-phase-error-formula}), (\ref{result^1st}), (\ref{result^2nd}), and (\ref{eq:realprob}) to obtain a relationship for the $k^{\rm th}$ pulse associated with the actual states: 
\begin{align}
0~& {\le} ~p_{Z_B} S_{+} {g^U} \left({\rm Tr}\big[{\hat P}(\ket{A_{+}}_{CB}){\hat M}^{(k)}_{+}\big],{1-\epsilon}\right)\nonumber\\
&-p_{Z_B} S_{-} {g^L} \left(\frac{p_{1_X}^{\rm vir}c}{S_{-}{p_{0_X}p_{X_B}}}P^{(k)}(q_{0x,0x}|{\rm Act})+\frac{1}{S_{-}p_{Z_B}}P^{(k)}({\rm ph|Act}),{1-\epsilon}\right),
\label{eq:refeq}
\end{align}
with ${\rm Tr}\big[{\hat P}(\ket{A_{+}}_{CB}){\hat M}^{(k)}_{+}\big]$ given by Eq.~(\ref{eq:part2}). We stress that, Eq.~(\ref{eq:refeq}) does not depend on ${\tilde p}^{\rm vir}_{\alpha_X}$, and the inner products $|{}_{CB}\langle A_{+}\ket{R_{+}}_{CB}|$ and $|{}_{CB}\langle A_{-}\ket{R_{-}}_{CB}|$ have the value of $1- \epsilon$. Therefore, Eq.~(\ref{eq:refeq}) does not depend on the inner products of the side-channel states nor on the inner products between the side-channel states and the qubit states. In particular, this means that our security proof works even if we do not know anything about the side-channel states and thus they can vary in time {and depend on the previous pulses,} as discussed above. Note that, Eq.~(\ref{eq:refeq}) is the required relationship for the actual states. This finishes the Deviation evaluation part.
\end{enumerate}

Finally, we have to convert Eq.~(\ref{eq:refeq}) into a relationship in terms of numbers, rather than probabilities. The procedure for this step is quite standard \cite{tamaki,mizutani,pereira,tamaki5}. {For this, first note that $g^{U}(x,y)$ and $-g^{L}(x,y)$ are concave functions with respect to $0\le x\le 1$ for any fixed $0\le y\le 1$. Also, recall that} the use of Azuma's inequality \cite{azuma} or Kato's inequality \cite{kato} converts the summed probabilities into the corresponding number in the asymptotic limit of a large number of pulses sent. That is, for $N\rightarrow \infty$, $\sum_{k}^{N}{P^{(k)}(q_{j,j_{B}}|{\rm Act})}\rightarrow {N(q_{j,j_{B}}|{\rm Act})}$ where ${N(q_{j,j_{B}}|{\rm Act})}$ is the number of events with Alice's setting choice equal to $j$ and Bob's outcome equal to $j_{B}$ in the experiment, after $N$ runs of the quantum communication protocol. Here, we emphasise that the use of Azuma's or Kato's inequality can deal with any correlations between Alice and Bob's measurement outcomes, making our proof valid against coherent attacks. Now, we take a summation over $k\in\{1,2,\cdots, N\}$ in Eq.~(\ref{eq:refeq}), and together with the two ingredients mentioned above we find the final expression as 
\begin{align} 
0& \le S_{+} {g^U} \left(\frac{p_{1_X}^{\rm vir}a}{S_{+} {p_{0_Z}p_{X_B}}}{\frac{N(q_{0z,0x}|{\rm Act})}{N}}+\frac{p_{1_X}^{\rm vir}b}{S_{+}{p_{1_Z}p_{X_B}}} {\frac{N(q_{1z,0x}|{\rm Act})}{N}}+\frac{p_{0_X}^{\rm vir}}{S_{+} {p_{0_X}p_{X_B}}}{\frac{N(q_{0x,1x}|{\rm Act})}{N}},{1-\epsilon}\right) \nonumber \\
&-S_{-} {g^L} \left(\frac{p_{1_X}^{\rm vir}c}{S_{-}{p_{0_X}p_{X_B}}}\frac{N(q_{0x,0x}|{\rm Act})}{{N}}+ \frac{1}{S_{-}p_{Z_B}} \frac{N(\rm {ph|Act})}{{N}}, {1-\epsilon}\right).
\label{eq:last}
\end{align}
Importantly, this inequality involves only the number of events defined in the actual protocol, and by solving this with respect to $N({\rm ph|Act})$, the security proof is done. We emphasise that our proof is valid for any coherent attack because {Eqs.~(\ref{eq:refeq}) and (\ref{eq:last}) hold} for any {${\hat K}_{\tilde e}^{(k)}$.}
 
\subsection{Arbitrarily long-range pulse correlations}
\label{sec:long_range_correlations}
In this section, we show how to extend our analysis to accommodate arbitrarily long-range correlations between the pulses. To simplify the  discussion, we consider the three-state protocol, but this formalism can be easily extended to any number of states. Our starting point is the assumption in Eq.~(\ref{eq:assumption}). We rewrite it here for convenience,
\begin{eqnarray}
\big|{}_{B_k} \braket{{\psi}_{j_k|j_{k-1},\cdots,j_{w+1},\tilde{j}_w,j_{w-1},\cdots,j_1} | {\psi}_{j_k|j_{k-1},\cdots,j_{w+1},j_w,j_{w-1},\cdots,j_1}} {}_{B_k}\big|^2&\geq&1-\epsilon_{k-w},
\label{eq:assumption3}
\end{eqnarray}
which holds for any $w$ $(1 \leq w \leq n)$ and $k$ $(w+1 \leq k \leq \min\{n,w+i\})$, where $i$ is the maximum correlation length. Note that the difference between both states is in the $j_{w}^{\text{th}}$ index. Also, the R.H.S. of Eq.~(\ref{eq:assumption3}) does not depend on the indices $j_k,j_{k-1},\cdots,j_1$ and $\tilde{j}_w$, and the term $k-w$ is associated with the correlation under consideration. For example, when $k-w = 1$ it refers to the nearest neighbour pulse correlation considered in the Results section. Furthermore, without loss of generality, we can assume the relation
\begin{eqnarray}
{}_{B_k}\braket{{\psi}_{j_k|j_{k-1},\cdots,j_{w+1},j_w,j_{w-1},\cdots,j_1}|{\psi}_{j_k|j_{k-1},\cdots,j_{w+1},0_X,j_{w-1},\cdots,j_1}} {}_{B_k} \geq 0,
\label{eq:assumption2}
\end{eqnarray}
after appropriately choosing the global phase of the state 
$\ket{\psi_{j_{k}|j_{k-1},...,j_{w+1},j_w,j_{w-1},...,j_1}}_{B_k}$ for any $w$ $(1 \leq w \leq n)$ and $k$ $(w+1 \leq k \leq \min \{n,w+i\})$. Using these assumptions, an extension of our framework is now presented. That is, we show how to obtain a lower bound on the parameter $a_j$ in Eq.~(\ref{eq:main}) starting from Eq.~(\ref{eq:assumption3}).

More generally, the entangled state prepared by Alice, shown in Eq.~(\ref{eq:entanglement}), can now be written as 
\begin{eqnarray}
\left|\Psi \right>_{AB}
&:=&
\sum_{j_n}\cdots\sum_{j_{1}}
\bigotimes_{\zeta=1}^n 
\ket{j_\zeta}_{A_\zeta}\ket{\psi_{j_\zeta|j_{\zeta-1},\cdots,j_1}}_{B_\zeta},
\label{eq:state_psi}
\end{eqnarray}
where $j_\zeta\in\{0_Z,1_Z,0_X\}$ and $\zeta \in \{1,2,\hdots,n\}$. Note that, $j_0$ represents having no condition, and the state $|\psi_{j_\zeta|j_{\zeta-1}, \cdots, j_1}\rangle_{B_\zeta}$ represents the long-range pulse correlations, that is, the state of the $\zeta^{\text{th}}$ pulse depends on all the previous setting choices. As before, we consider the $t^{\rm th}$ virtual protocol, where Alice measures her ancilla systems up to the $k^{\text{th}}$ pulse. More precisely, she measures the key generation rounds with tag $t$ by using the complementary basis, and measures all the other $k-1$ rounds as in the actual protocol. The whole (unnormalised) state can then be expressed as
\begin{align}
\ket{\Psi_{j'_{k-1},\cdots,j'_1}}_{AB}
&:= \Bigg(\bigotimes_{\tilde{\zeta}=1}^{k-1} \ket{j'_{\tilde\zeta}}_{A_{\tilde\zeta}} \ket{\psi_{j'_{\tilde\zeta}|j'_{\tilde\zeta-1},\cdots,j'_1}}_{B_{\tilde\zeta}}\Bigg) \otimes \sum_{j_k} \ket{j_k}_{A_k} \ket{\psi_{j_k|j'_{k-1},\cdots,j'_1}}_{B_k} \nonumber \\
&\otimes \Bigg(\sum_{j_n}\cdots\sum_{j_{k+1}}
\bigotimes_{\zeta=k+1}^n 
\ket{j_\zeta}_{A_\zeta}\ket{\psi_{j_\zeta|j_{\zeta-1},\cdots,j_{k+1},j_{k},j'_{k-1}, \cdots,j'_1}}_{B_\zeta}\Bigg).
\label{eq:state_psi2}
\end{align}
To clarify, after Alice's measurement, the state $\ket{\Psi}_{AB}$ in Eq.~(\ref{eq:state_psi}) becomes the state $\ket{\Psi_{j'_{k-1},\cdots,j'_1}}_{AB}$ in Eq.~(\ref{eq:state_psi2}), where the subscripts indicate its dependence on the previous measurement results $j_{k-1}',\cdots,j_1'$. Note that, Eq.~(\ref{eq:state_psi2}) corresponds to Eq.~(\ref{eq:protocol3}) in the Results section.

Now, similarly to our analysis for the nearest neighbour pulse correlations, in order to see how the information $j_{k}$ is encoded in the state $\ket{\Psi_{j'_{k-1},\cdots,j'_1}}_{AB}$, defined in Eq.~(\ref{eq:state_psi2}), we rewrite it as
\begin{eqnarray}
&&\ket{\Psi_{j'_{k-1},\cdots,j'_1}}_{AB} = \Bigg(\bigotimes_{\tilde{\zeta}=1}^{k-1} \ket{j'_{\tilde\zeta}}_{A_{\tilde\zeta}} \ket{\psi_{j'_{\tilde\zeta}|j'_{\tilde\zeta-1},\cdots,j'_1}}_{B_{\tilde\zeta}}\Bigg)
\otimes \sum_{j_k} \ket{j_k}_{A_k} \ket{\psi_{j_k|j'_{k-1},\cdots,j'_1}}_{B_k}
\nonumber\\
&&\otimes \left(a_{j_k,j'_{k-1},\cdots,j'_1}
\ket{\Phi_{j'_{k-1},\cdots,j'_1}}_{A_{k+1}, \cdots, A_n,B_{k+1},\cdots, B_{n}} +
b_{j_k,j'_{k-1},\cdots,j'_1}
\ket{\Phi^\perp_{j_k,j'_{k-1},\cdots,j'_1}}_{A_{k+1}, \cdots, A_n,B_{k+1},\cdots,B_{n}}
\right),
\label{eq:state_psi3}
\end{eqnarray}
where $\ket{\Phi_{j'_{k-1},\cdots,j'_1}}_{A_{k+1}, \cdots, A_n,B_{k+1},\cdots, B_{n}}$, and $\ket{\Phi^\perp_{j_k,j'_{k-1},\cdots,j'_1}}_{A_{k+1}, \cdots, A_n,B_{k+1},\cdots,B_{n}}$ are normalised states, and 
$\ket{\Phi_{j'_{k-1},\cdots,j'_1}}_{A_{k+1}, \cdots, A_n,B_{k+1},\cdots, B_{n}}$ is orthogonal to $\ket{\Phi^\perp_{j_k,j'_{k-1},\cdots,j'_1}}_{A_{k+1}, \cdots, A_n,B_{k+1},\cdots,B_{n}}$. Recall that, the subscripts in the variables, e.g. $a_{j_k,j'_{k-1},\cdots,j'_1}$, $b_{j_k,j'_{k-1},\cdots,j'_1}$, or in the state $\ket{\Psi_{j'_{k-1},\cdots,j'_1}}_{AB}$, indicate their dependence on previous results. Importantly, the state $\ket{\Phi_{j'_{k-1},\cdots,j'_1}}_{A_{k+1}, \cdots, A_n,B_{k+1},\cdots, B_{n}}$ does not depend on $j_{k}$ but $\ket{\Phi^\perp_{j_k,j'_{k-1},\cdots,j'_1}}_{A_{k+1}, \cdots, A_n,B_{k+1},\cdots,B_{n}}$ does. In other words, $\ket{\Phi^\perp_{j_k,j'_{k-1},\cdots,j'_1}}_{A_{k+1}, \cdots, A_n,B_{k+1},\cdots,B_{n}}$ is the side-channel information for $j_{k}$. Furthermore, note that $\ket{\psi_{j_{k}|j'_{k-1},\cdots,j'_1}}_{B_{k}} \otimes \ket{\Phi^\perp_{j_k,j'_{k-1},\cdots,j'_1}}_{A_{k+1}, \cdots, A_n,B_{k+1},\cdots,B_{n}}$ in Eq.~(\ref{eq:state_psi3}) corresponds to $\ket{{\phi}_{j_k|j'_{k-1}}^\perp}_{A_{k+1},\cdots,A_n,B_k,B_{k+1},\cdots,B_n}$ in Eq.~(\ref{loss-tolerant-side-channel}).

Next, we obtain a lower bound on the coefficient $a_{j_k,j'_{k-1}, \cdots, j'_1}$. For $\ket{\Phi_{j'_{k-1},\cdots,j'_1}}_{A_{k+1}, \cdots, A_n,B_{k+1},\cdots, B_{n}}$, one may choose a state such that it becomes independent of $j_{k}$. One of such choices could be
\begin{eqnarray}
\ket{\Phi_{j'_{k-1},\cdots,j'_1}}_{A_{k+1}, \cdots, A_n,B_{k+1},\cdots, B_{n}}:= \sum_{j_n}\cdots\sum_{j_{k+1}}
\bigotimes_{\zeta=k+1}^n
\ket{j_\zeta}_{A_\zeta}\ket{{\psi}_{j_\zeta|j_{\zeta-1},\cdots,j_{k+1},0_X,j'_{k-1},\cdots,j'_1}}_{B_\zeta},
\label{eq:vector}
\end{eqnarray}
which is the state of the last $(n-k)$ systems in Eq.~(\ref{eq:state_psi2}) with only the $k^{\text{th}}$ index of $\ket{{\psi}_{j_\zeta|j_{\zeta-1},\cdots,j_{k+1},j_k,j'_{k-1},\cdots,j'_1}}_{B_\zeta}$ being fixed to $0_X$. Importantly, this state is independent of $j_k$. Since $a_{j_k,j'_{k-1},\cdots,j'_1}$ is equal to the inner product between the state given by Eq.~(\ref{eq:vector}) and the vector
\begin{eqnarray}
\sum_{j_n}\cdots\sum_{j_{k+1}}
\bigotimes_{\zeta=k+1}^n 
\ket{j_\zeta}_{A_\zeta}\ket{\psi_{j_\zeta|j_{\zeta-1},\cdots,j_{k+1},j_{k},j'_{k-1},\cdots,j'_1}}_{B_\zeta},
\end{eqnarray}
which is the expression in the last parenthesis of Eq.~(\ref{eq:state_psi2}), we can evaluate a lower bound for $a_{j_k,j'_{k-1},\cdots,j'_1}$ as
\begin{eqnarray}
|a_{j_k,j'_{k-1},\cdots,j'_1}|&&= ~\bigg|\sum_{j_n}\cdots\sum_{j_{k+1}}
\prod_{\zeta=k+1}^n p_{j_\zeta} ~{}_{B_\zeta}\braket{{\psi}_{j_\zeta|j_{\zeta-1},\cdots,j_{k+1},0_X,j'_{k-1},\cdots,j'_1} | {\psi}_{j_\zeta|j_{\zeta-1},\cdots,j_{k+1},j_{k},j'_{k-1},\cdots,j'_1}} {}_{B_\zeta}\bigg| \nonumber \\
&& =\sum_{j_n}\cdots\sum_{j_{k+1}}
\prod_{\zeta=k+1}^n p_{j_\zeta} ~
\big| {}_{B_\zeta}\braket{{\psi}_{j_\zeta|j_{\zeta-1},\cdots,j_{k+1},0_X,j'_{k-1},\cdots,j'_1} | {\psi}_{j_\zeta|j_{\zeta-1},\cdots,j_{k+1},j_{k},j'_{k-1},\cdots,j'_1}} {}_{B_\zeta} \big| \nonumber\\
&&\geq \sum_{j_{k+i}}\cdots\sum_{j_{k+1}}
\prod_{\zeta=k+1}^{k+i} p_{j_\zeta} (1-\epsilon_{\zeta-k})^{1/2} \nonumber \\
&&= \prod_{\zeta=1}^{i} (1-\epsilon_\zeta)^{1/2}.
\end{eqnarray}
In the second equality we use the result given by Eq.~(\ref{eq:assumption2}) and the inequality comes from Eq.~(\ref{eq:assumption3}).

\section{Data availability}
All data needed to evaluate the conclusions in the paper are present in the paper and/or the Supplementary Material. Additional data available from authors upon request.

\section{Acknowledgements}
We thank Guillermo Currás-Lorenzo and Álvaro Navarrete for very valuable discussions. We also sincerely thank an anonymous referee for finding a mistake in a previous version of this manuscript. This work was supported by the Spanish Ministry of Economy and Competitiveness (MINECO), the Fondo Europeo de Desarrollo Regional (FEDER) through the grant TEC2017-88243-R, and the European Union's Horizon 2020 research and innovation programme under the Marie Sk\l{}odowska-Curie grant agreement No 675662 (project QCALL). K.T. acknowledges support from JSPS KAKENHI Grant Numbers JP18H05237 18H05237 and JST-CREST JPMJCR 1671.

%\section{Competing interests}
%The authors declare no competing interests. 

\section{Author contributions}
M.P. and K.T. conceived the initial idea of how to deal with pulse correlations. {K.T. constructed the reference technique through discussions with all the other authors.} Then, G.K., A.M., and M.C. contributed to the development {and discussion} of the initial idea and {the reference technique} with M.P. and K.T.. M.P performed the numerical simulations. All authors contributed to writing the manuscript. \newpage

\quad \quad 
\begin{center}
{\large{\textbf{SUPPLEMENTARY MATERIAL}}}
\end{center} 

%%%%%%%%%%%%%%%%%
\setcounter{equation}{0}
\setcounter{figure}{0}
\setcounter{table}{0}
\setcounter{section}{0}
\makeatletter
\renewcommand{\theequation}{S\arabic{equation}}
\renewcommand{\thefigure}{S\arabic{figure}}
\renewcommand{\thesection}{S\arabic{section}}

\section{Bound on the reference states}
The RT introduced in this work requires to quantify the maximum deviation between the probabilities associated with the actual states and \textcolor{black}{those} associated with the reference states. In this section, we derive a bound to evaluate this difference. Note that, the bound obtained is essentially equivalent to \textcolor{black}{that} given in Eq.~(A.12) of the Lo-Preskill's \textcolor{black}{(LP)} analysis introduced in [20]. However, the derivation presented below is simpler since it is a purely mathematical inequality which does not use the physical intuition \textcolor{black}{involving} the quantum coin idea.

One wants to obtain an inequality of the following form
\begin{eqnarray}
|\braket{A|R}| \le \sqrt{\bra{A} \hat{M} \ket{A} \bra{R} \hat{M} \ket{R}} +  \sqrt{\bra{A} (\hat{I} - \hat{M}) \ket{A} \bra{R} (\hat{I} - \hat{M}) \ket{R}},
\label{eq:ineq}
\end{eqnarray} 
where $\ket{A}$ and $\ket{R}$ are any normalised states associated with the actual and reference states respectively, and $\hat{M}$ is any measurement operator such that $0 \le \hat{M} \le 1$. To guarantee this in practice, Alice and Bob must run the quantum key distribution (QKD) protocol sequentially, i.e.~Alice only emits the next pulse after Bob has measured the previous one. The proof of Eq.~(\ref{eq:ineq}) is as follows.

Consider two vectors $\sqrt{\hat{N}} \ket{A}$ and $\sqrt{\hat{N}} \ket{R}$ for any $0 \le \hat{N}$. Then, by applying \textcolor{black}{the} Cauchy-Schwarz inequality to \textcolor{black}{these} two vectors we have that 
\begin{eqnarray}
|\bra{A}\hat{N}\ket{R}| \le \sqrt{\bra{A}\hat{N}\ket{A} \bra{R}\hat{N}\ket{R}}.
\label{eq:ineq2}
\end{eqnarray}
Since the real part of a complex number is equal to or smaller than its absolute value, one can re-write Eq.~(\ref{eq:ineq2}) as
\begin{eqnarray}
\frac{1}{2} \left(\bra{A}\hat{N}\ket{R} + \bra{R}\hat{N}\ket{A} \right) \le \sqrt{\bra{A}\hat{N}\ket{A} \bra{R}\hat{N}\ket{R}}.
\label{eq:ineq3}
\end{eqnarray}
Next, consider two inequalities in the form of Eq.~(\ref{eq:ineq3}) such that in one inequality we set $\hat{N} = \hat{M}$ and in the other $\hat{N} = (\hat{I}-\hat{M})$. By adding these inequalities, we obtain
\begin{eqnarray}
\frac{1}{2} \left(\braket{A|R} + \braket{R|A} \right) \le \sqrt{\bra{A} \hat{M}\ket{A} \bra{R} \hat{M}\ket{R}} + \sqrt{\bra{A}(\hat{I}-\hat{M})\ket{A} \bra{R}(\hat{I}-\hat{M})\ket{R}}.
\end{eqnarray} 
Note that, one has the freedom to choose the \textcolor{black}{global} phase of the quantum state $\ket{R}$ such that the inner product $\braket{A|R}$ becomes real and positive. By exploiting this freedom, we have that 
\begin{eqnarray}
|\braket{A|R}| \le \sqrt{\bra{A} \hat{M}\ket{A} \bra{R} \hat{M}\ket{R}} + \sqrt{\bra{A}(\hat{I}-\hat{M})\ket{A} \bra{R}(\hat{I}-\hat{M})\ket{R}},
\end{eqnarray} 
for any two quantum states $\ket{A}$ and $\ket{R}$, thus obtaining the desired inequality stated in Eq.~(\ref{eq:ineq}).

Alternatively, this inequality can be written as 
\begin{eqnarray}
|\braket{A|R}| \le \sqrt{\Tr \big[\dyad{A}{A} \hat{M}\big]\Tr \big[\dyad{R}{R} \hat{M}\big]} + \sqrt{\left(1-\Tr \big[\dyad{A}{A} \hat{M}\big]\right)\left(1-\Tr \big[\dyad{R}{R} \hat{M}\big]\right)}. 
\label{eq:Bref}
\end{eqnarray}
By solving \textcolor{black}{Eq.~(\ref{eq:Bref})} with respect to $\Tr \big[\dyad{R}{R} \hat{M}\big]$ we obtain 
\textcolor{black}{
\begin{eqnarray}
g^L\left(\Tr \big[\dyad{A}{A} \hat{M}\big],|\braket{A|R}| \right) \le \Tr \big[\dyad{R}{R} \hat{M}\big] \le g^U\left(\Tr \big[\dyad{A}{A} \hat{M}\big],|\braket{A|R}|\right),
\label{eq:bound}
\end{eqnarray} where \begin{align}
g^L(x,y) = \left\{
        \begin{array}{ll}
            0 & \quad x  < 1 - y^2  \\
            x + (1-y^2)(1-2x) - 2y\sqrt{(1-y^2)x(1-x)} & \quad x  \geq 1 - y^2, 
        \end{array}
    \right. 
\end{align}
and
\begin{align}
g^U(x,y) = \left\{
        \begin{array}{ll}
            x + (1-y^2)(1-2x) + 2y\sqrt{(1-y^2)x(1-x)} & \quad x \leq y^2  \\
            1 & \quad x > y^2.
        \end{array}
    \right.
\end{align}}Note that, this bound can be written in terms of the actual yields obtained from a practical implementation of the protocol and the selected reference and actual states.

\section{GLLP type security proofs as a special case of the reference technique}
\subsection{\textcolor{black}{Derivation of the GLLP formula from the reference technique}}
The RT can be seen as a generalisation of the loss-tolerant (LT) protocol for qubit states, and of the GLLP type security proofs for non-qubit states [19-21]. In fact, as we have already shown in the Main text, the RT includes the LT protocol in the Reference formula part. \textcolor{black}{Here}, we reconstruct the GLLP type security proof starting from the RT. More specifically, we prove the security of a four-state protocol in which Alice employs the state $\ket{\psi_{1_X}}_{B}$ besides the three states $\ket{\psi_{0_Z}}_{B}$, $\ket{\psi_{1_Z}}_{B}$, and $\ket{\psi_{0_X}}_{B}$, and for simplicity, we assume that Alice chooses these states randomly for each pulse emission. We remark that, no further assumptions on these states are required. That is, they can be qubit states or linearly independent states. For simplicity of discussion, here we consider the absence of pulse correlations. As explained in the Main text, however, the inclusion of this imperfection would simply result in the consideration of $(i+1)$ virtual protocols, where $i$ is the maximum correlation length, and the application of the RT to each of them. As for Bob, we assume that he selects the $Z$ basis or the $X$ basis with probabilities $p_{Z_{B}}$ and $p_{X_{B}}$, respectively, \textcolor{black}{and the basis independent efficiency condition is satisfied.}

From the perspective of the RT, the GLLP security proof [19] and other related analyses, such as, \textcolor{black}{for instance,} the LP analysis [20] and \textcolor{black}{that} in [21], skip the Reference formula part and directly \textcolor{black}{apply} the Deviation evaluation part. To see this, first recall that the required quantity to prove the security of a QKD protocol is the number \textcolor{black}{of} phase errors (however, for simplicity of discussion we consider the probability of a phase error). As explained in the Main text, the phase errors are defined in the $X$ basis and they correspond to the errors that Alice and Bob would have obtained if they had performed an $X$-basis measurement on the $Z$-basis states defined in \textcolor{black}{Eq.~(28)}. The probability of a phase error  for the $k^{\rm th}$ pulse when employing the actual states can then be defined as
\begin{eqnarray}
P^{(k)}({\rm ph|Act}):=p_{Z_A}p_{Z_B}{\rm Tr}\big[{\hat P}(\ket{A}_{CB}) {\hat M}^{(k)}_{\rm ph}\big],
\label{eq:ephGLLP}
\end{eqnarray}
with 
\begin{align}
&\ket{A}_{CB}=\frac{1}{\sqrt{2}}\ket{0_{z,A}, Z_{B}}_{C}\ket{\psi_{0_Z}}_{B}+\frac{1}{\sqrt{2}}\ket{1_{z,A}, Z_{B}}_{C}\ket{\psi_{1_Z}}_{B},\nonumber\\
&{\hat M}^{(k)}_{\rm ph}:={\hat P}\left(\frac{\ket{0_{z,A}, Z_{B}}_{C}-\ket{1_{z,A}, Z_{B}}_{C}}{\sqrt{2}}\right)\otimes{\hat M}_{0_X}^{(k)}+{\hat P}\left(\frac{\ket{0_{z,A}, Z_{B}}_{C}+\ket{1_{z,A}, Z_{B}}_{C}}{\sqrt{2}}\right)\otimes{\hat M}_{1_X}^{(k)},
\label{eq:a&m}
\end{align}
where $\hat{P}(\ket{\cdot}) = \dyad{\cdot}{\cdot}$. \textcolor{black}{In Eq.~(\ref{eq:a&m}),} we \textcolor{black}{assume} that Alice chooses a bit value at random when she selects the $Z$ basis in the protocol. Also, note that $\left\{{\hat P}\left(\frac{\ket{0_{z,A}, Z_{B}}_{C}-\ket{1_{z,A}, Z_{B}}_{C}}{\sqrt{2}}\right),{\hat P}\left(\frac{\ket{0_{z,A}, Z_{B}}_{C}+\ket{1_{z,A}, Z_{B}}_{C}}{\sqrt{2}}\right)\right \}$ is the basis for Alice's $X$-basis projection measurement.  \textcolor{black}{This means that}, ${\hat M}^{(k)}_{\rm ph}$ is the positive-operator-valued-measure (POVM) element that represents the occurrence of the phase errors after Eve's intervention. As before, in order to estimate the probability of a phase error, we require the reference states. In the GLLP type security analyses, the actual $X$-basis states serve as the reference states, and we may select
\begin{eqnarray}
\ket{R}_{CB}=\frac{1}{\sqrt{2}}\left(\frac{\ket{0_{z,A}, Z_{B}}_{C}+\ket{1_{z,A}, Z_{B}}_{C}}{\sqrt{2}}\right)\ket{\psi_{0_X}}_{B}+ \frac{1}{\sqrt{2}}\left(\frac{\ket{0_{z,A}, Z_{B}}_{C}-\ket{1_{z,A}, Z_{B}}_{C}}{\sqrt{2}}\right)\ket{\psi_{1_X}}_{B},
\end{eqnarray}
where we \textcolor{black}{assume} that Alice chooses a bit value at random when she selects the $X$ basis in the protocol. For the transformation of Eq.~(\ref{eq:ephGLLP}) into a relationship associated with the reference states, we use \textcolor{black}{Eq.~(\ref{eq:bound})}. Note that, in this case the roles of $\ket{R}$ and $\ket{A}$ are switched. Finally, we obtain the equation essentially presented in [19-21] as
\begin{align}
P^{(k)}({\rm{ph|Act}}) \le p_{Z_A}p_{Z_B} \textcolor{black}{g^U}\left(\frac{P^{(k)}(\rm{X-error|Act})}{p_{X_A}p_{X_B}}, |{}_{CB}\langle R\ket{A}_{CB}|\right),
\label{GLLP}
\end{align}where
\begin{align}
P^{(k)}({\rm X-error|Act}):=& ~p_{X_A}p_{X_B}{\rm Tr}\big[{\hat P}(\ket{R}_{CB}) {\hat M}^{(k)}_{\rm ph}\big]\nonumber\\
=& ~p_{1_X}p_{X_B}{\rm Tr}\big[{\hat P}(\ket{\psi_{1_X}}_{B}) {\hat M}^{(k)}_{0_X}\big]+p_{0_X}p_{X_B}{\rm Tr}\big[{\hat P}(\ket{\psi_{0_X}}_{B}) {\hat M}^{(k)}_{1_X}\big],
\label{Lo-Preskill-reproduced}
\end{align}
is the probability that Alice and Bob select the $X$ basis and they observe a bit error in the actual protocol. This means that, we can estimate the probability of a phase error from a quantity that is directly \textcolor{black}{observed in the} experiment, and therefore guarantee the security of the QKD protocol. Note that, if we take the states $\{\ket{\psi_j}\}_j$ as those in Eq.~(1), the calculation of $|{}_{CB}\langle R\ket{A}_{CB}|$ in Eq.~(\ref{Lo-Preskill-reproduced}) involves the evaluation of the inner products of the unknown side-channel states (i.e.~$|{}_{B}\langle \phi^{\perp}_j \ket{\phi^{\perp}_{\tilde j}}{}_{B}|$ and $|{}_{B}\langle \phi^{\perp}_j \ket{\phi_{\tilde j}}{}_{B}|$ for $j\neq \tilde j$).

Now, let us discuss the consequences of skipping the Reference formula part of the RT. First, suppose that all the states are qubit states. According to the LP analysis [20], the resulting secret key rate based on Eq.~(\ref{Lo-Preskill-reproduced}) for the qubit-based four-state protocol is \textit{not} tolerant to channel loss when there are state preparation flaws (SPFs), i.e.~{$|{}_{CB}\langle R\ket{A}_{CB}| \ne 1$}. This is due to a potential enhancement of the imperfection, represented by $1-\textcolor{black}{|{}_{CB}\langle R\ket{A}_{CB}|}$, with channel loss. From the viewpoint of the RT, we can attribute this enhancement to the fact that the analysis lacks the use of the Reference formula, from which the qubit-based protocol becomes loss-tolerant for a proper choice of \textcolor{black}{the} reference states. \textcolor{black}{In what follows, we evaluate the secret key rate and demonstrate the consequences of skipping the Reference formula part.  }
\subsection{\textcolor{black}{Simulation of the secret key rate}}
\textcolor{black}{In this section, we compare the RT based on the original LT protocol, introduced in the Main text, with the RT based on the GLLP type security proofs. It is important to note \textcolor{black}{however} that, the comparison between these \textcolor{black}{two} cases of the RT might be considered unfair for the following reasons. The RT based on the GLLP type security proofs requires four states, instead of three states, and 
%It has been shown that when there are no source imperfections, using four states does not result in higher secret key rates. In fact, in \cite{tamaki} the authors demonstrate that a three-state protocol achieves the same performance as a four-state protocol since both protocols can obtain an exact value for the phase error rate. However, this is not the case in the presence of source imperfections. That is, depending on the protocol and consequently on the security analysis performed, different estimations of the phase error rate are obtained. For instance, by comparing Eq.~(24) and Eq.~(\ref{GLLP}) we can see that when using a three-state protocol the bound in Eq.~(\ref{eq:bound}) needs to be used twice, however, for a four-state protocol it is only used once. In turn, Eq.~(\ref{GLLP}) results in a tighter estimation of the phase error rate in certain parameter regimes. 
\textcolor{black}{it} requires analytical or numerical optimisation. This is \textcolor{black}{so} because the inner product $|{}_{CB}\langle R\ket{A}_{CB}|$ in Eq.~(\ref{GLLP}) depends on the inner products of the side-channel states, i.e.~the terms ${}_{B}\langle \phi_j^\perp\ket{\phi_{\tilde{j}}^\perp} {}_{B}$ and ${}_{B}\langle \phi_j^\perp\ket{\phi_{\tilde{j}}} {}_{B}$ for $j \neq \tilde{j}$. Since these side-channel states are unknown, the values of their inner products need to be optimised such that the secret key rate is minimised. In the RT based on the original LT protocol, \textcolor{black}{however,} this optimisation is not required since \textcolor{black}{it is} fully analytical even when the side-channel states are totally unknown (see the Results section). }

For simplicity, in the simulations we assume that the inner product ${}_{B}\langle \phi_j^\perp\ket{\phi_{\tilde{j}}} {}_{B} = 0$ for all $j$ \textcolor{black}{and} $\tilde{j}$. Note that, this case does not correspond to the nearest neighbour pulse correlations described in Eq.~(10). That is, in Eq.~(10), the side-channel states $\ket{\phi_j^\perp}_B$ live in the same qubit space as the states $\ket{\phi_j}_B$, and therefore ${}_{B}\langle \phi_j^\perp\ket{\phi_{\tilde{j}}} {}_{B} \ne 0$ for all $j$ \textcolor{black}{and} $\tilde{j}$, given the two-dimensionality of the space. The scenario evaluated in this \textcolor{black}{subsection} corresponds to the case when $\ket{\phi_j^\perp}_B$ lives in a higher dimensional Hilbert space. Examples of this scenario include THAs, mode dependencies or a combination of the two. As explained in the Main text, all these imperfections can be accommodated simultaneously through the parameter $a_j$ in Eq.~(1).

\begin{figure}[h]
	\centering
	 \subfloat[]{\includegraphics[width=0.4\columnwidth]{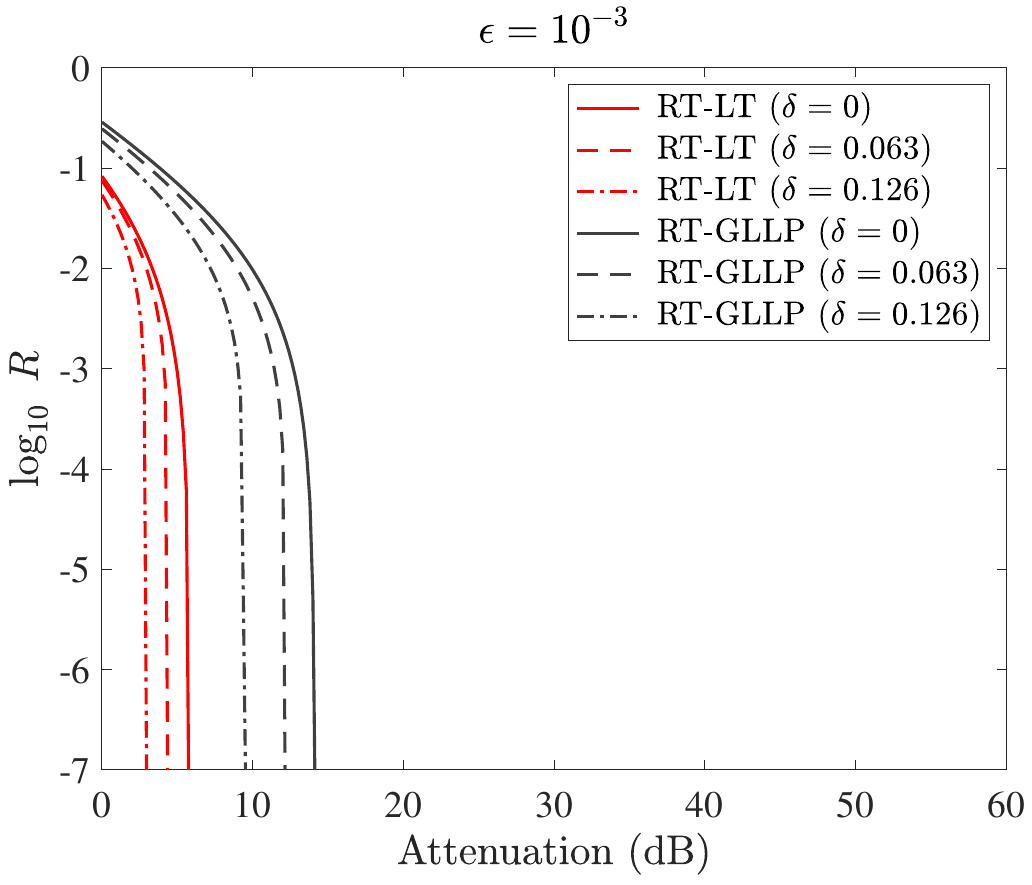}}
     ~~~
    \subfloat[]
    {\includegraphics[width=0.4\columnwidth]{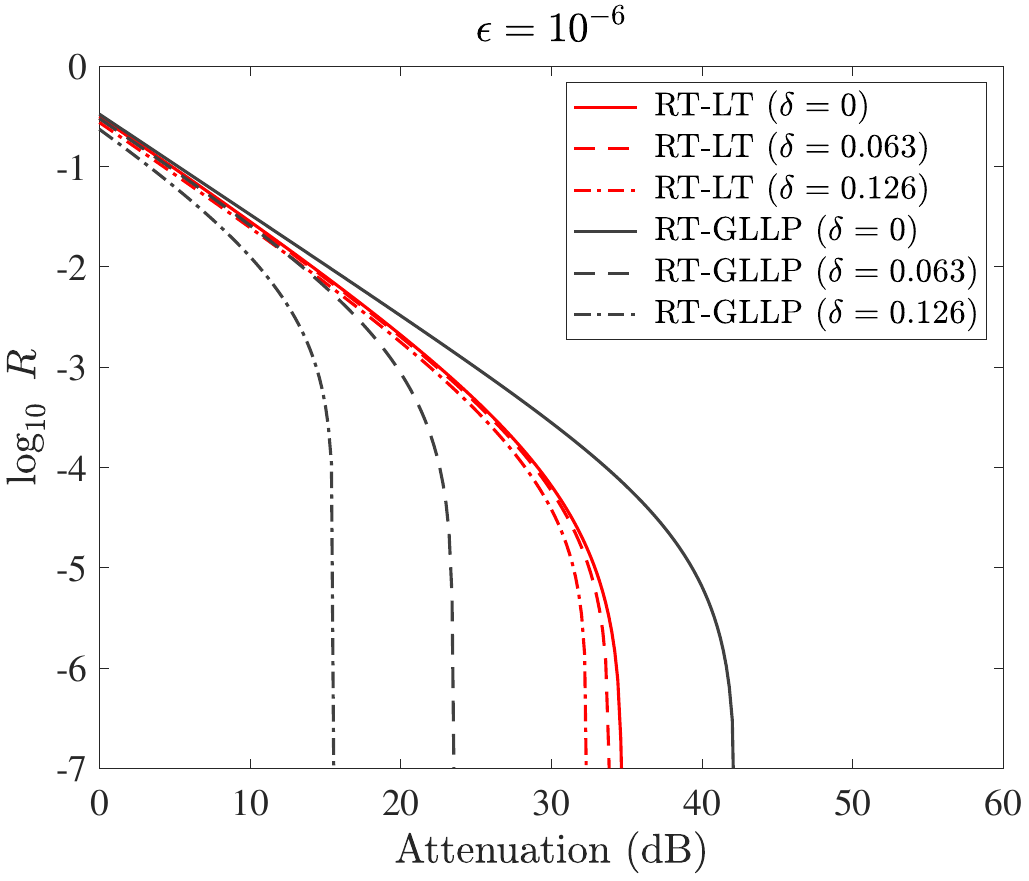}}
	\caption{Secret key rate $R$ against the overall system loss \textcolor{black}{measured} in dB in the presence of imperfections when using the reference technique based on the original loss-tolerant (RT-LT) protocol and the reference technique based on the GLLP type security proofs (RT-GLLP). In all graphs, the \textcolor{black}{red} and the black lines are associated with the RT-LT and the RT-GLLP, respectively. The solid lines correspond to $\delta=0$, while the dashed (dashed-dotted) lines correspond to $\delta=0.063$ ($\delta = 0.126$) as indicated in the legend. (a) When $\epsilon$ is large, the RT-GLLP outperforms the RT-LT for any value of $\delta$. (b) As $\epsilon$ decreases, \textcolor{black}{both} security proofs provide higher secret key rates but while the RT-GLLP drastically decreases as the SPFs increase, the RT-LT maintains \textcolor{black}{its} performance.}
	\label{fig:comparison2}
\end{figure}

\textcolor{black}{For the simulations of the RT based on the GLLP type security proofs, we use the three states $\ket{\psi_{0_Z}}_{B}$, $\ket{\psi_{1_Z}}_{B}$, and $\ket{\psi_{0_X}}_{B}$, defined in the Eq.~(27), and \textcolor{black}{we assume that} the fourth state \textcolor{black}{$\ket{\psi_{1_X}}_B$, is given by} 
\begin{equation}
\textcolor{black}{\ket{\psi_{1_X}}_B = (1-\epsilon) \ket{\phi_{1_X}}_B + \sqrt{1-(1-\epsilon)^2} \ket{\phi_{1_X}^\perp}_B},
\end{equation}
where $\epsilon$ represents the deviation from the ideal qubit scenario due to imperfections, such as THAs and mode dependencies. Here, the qubit state \textcolor{black}{$\ket{\phi_{1_X}}_B$} is defined as in [18] such that} 
\begin{equation}
\textcolor{black}{\ket{\phi_{1_X}}_B} = \cos(\frac{3\pi}{4} + \frac{3 \delta}{4}) \ket{0_Z}_B + \sin(\frac{3\pi}{4} + \frac{3 \delta}{4}) \ket{1_Z}_B,
\end{equation}
\noindent  and \textcolor{black}{$\ket{\phi_{1_X}^\perp}_B$} is a state orthogonal to \textcolor{black}{the qubit space.} Also, we use the same experimental parameters and the same channel model as in the Main text, and, for simplicity, we assume that the probabilities for Alice and Bob to select the $Z$ basis are $p_{Z_A} = \frac{2}{3}$ and $p_{Z_B} = \frac{1}{2}$, respectively. To evaluate the source imperfections we select the values $\epsilon = 10^{-3}$ and $\epsilon = 10^{-6}$. Regarding the SPFs, we choose $\delta = 0$, $\delta = 0.063$ and $\delta = 0.126$ according to the experimental results reported in [43-45]. Note that, here we choose a larger range of $\delta$ than in the Main text because we want to clearly \textcolor{black}{show} the effect that the SPFs have on the secret key rate \textcolor{black}{with channel loss}. The results are shown in Fig. \ref{fig:comparison2}.

%\begin{figure}[h]
%	\begin{subfigure}[H]{0.45\textwidth}
%	\includegraphics[width=8cm]{comp3_final} 
%	\caption{}
%	\end{subfigure}
%	\begin{subfigure}[H]{0.45\textwidth}
%	\includegraphics[width=8cm]{comp4_final}
%	\caption{}
%	\end{subfigure}
%	\caption{\textcolor{black}{Secret key rate $R$ against the overall system loss \textcolor{black}{measured} in dB in the presence of imperfections when using the reference technique based on the original loss-tolerant (RT-LT) protocol and the reference technique based on the GLLP type security proofs (RT-GLLP). In all graphs, the \textcolor{black}{red} and the black lines are associated with the RT-LT and the RT-GLLP, respectively. The solid lines correspond to $\delta=0$, while the dashed (dashed-dotted) lines correspond to $\delta=0.063$ ($\delta = 0.126$) as indicated in the legend. (a) When $\epsilon$ is large, the RT-GLLP outperforms the RT-LT for any value of $\delta$. (b) As $\epsilon$ decreases, \textcolor{black}{both} security proofs provide higher secret key rates but while the RT-GLLP drastically decreases as the SPFs increase, the RT-LT maintains \textcolor{black}{its} performance.} }
%	\label{fig:comparison2}
%\end{figure}

\textcolor{black}{A striking difference between the RT based on the original LT protocol and the RT based on the GLLP type security proofs is that the latter skips the Reference formula part of the RT. Therefore, it is expected that the secret key rate for the RT based on the GLLP type security proofs is more vulnerable to SPFs, especially when $\epsilon$ is small. In fact, Fig. S1(b) clearly shows this tendency. On the other hand, when $\epsilon$ increases, the RT based on the GLLP security proofs outperforms the RT based on the original LT protocol. This can also be seen by comparing Eq.~(24) and Eq.~(S13). That is, when using a three-state protocol the bound in Eq.~(S9) needs to be used twice, however, for}
\noindent \textcolor{black}{a four-state protocol it is only used once. In turn, Eq.~(S13) results in a tighter estimation of
the phase error rate in certain parameter regimes, especially when $\epsilon$ is large.}

\textcolor{black}{Fig. S1 suggests that depending on the parameter regimes for $\epsilon$ and SPFs, one should \textcolor{black}{select the best approach} for the security proof. However, \textcolor{black}{we note that} this could be circumvented by considering a security proof for a four-state protocol, which accommodates the Reference formula part (like the RT-LT) and uses the bound in Eq. (S9) only once (like the RT-GLLP). This way, the resulting key rate is expected to be robust against an increase of both SPFs and $\epsilon$. We leave its explicit analysis and further exploration of the RT for future works.}

%\textcolor{black}{The results presented in Fig.~\ref{fig:comparison2}(a) might lead us to believe that the RT based on the GLLP security proofs results in a better performance in general. We remark again that, this comparison is unfair because the RT based on the GLLP security proofs requires four states as well as numerical optimisation. Finally, we emphasise that the RT could be used with a four-state protocol without skipping the Reference formula part, resulting in a security analysis that could be robust against most typical device imperfections. More precisely, it would inherit from the original LT protocol high tolerance against SPFs, and by considering a four-state protocol we would be able to obtain a tighter estimation of the phase error rate since the bound in Eq.~(\ref{eq:bound}) would only be used once. However, we leave this analysis and further exploration of the RT for future works.  }

\section{Sufficient conditions to apply the reference technique}
\label{Sufficient conditions for the RT}
In this section, we discuss the sufficient conditions required to apply the RT to prove the security of a QKD protocol. To be precise, we consider an actual $m$-state protocol where Alice chooses the setting $j$ with probability $p_j$, and sends the normalised state $\ket{\psi_j}_{B}$. Note that here we do not consider a particular source imperfection and the normalised states in $\{\ket{\psi_j}_{B}\}_{j=1, 2,\cdots, m}$ can be qubit or non-qubit states. However, as explained in the Main text, in the presence of pulse correlations, with maximum correlation length $i$, the RT would need to be applied separately to each of the $(i+1)$ virtual protocols. Moreover, we assume that Bob performs a measurement, which satisfies the basis independent detection efficiency condition, and we denote the set of its outcomes by $\{\gamma\}_{\gamma=1,2,\cdots}$. In order to prove the security of this $m$-state protocol we apply the RT as follows. \\

\begin{enumerate}
\item \underline{{Reference formula part}}

First, we consider an entanglement-based protocol for the actual protocol, and we assume that we can define the events $q_{es}$ and \textcolor{black}{$q_{obs_{j,\gamma}}$}. Here, $q_{es}$ represents the event associated with the occurrence of a quantity that we \textcolor{black}{wish} to estimate in the security proof, such as the phase errors, and \textcolor{black}{$q_{obs_{j,\gamma}}$} represents the actual event in which Alice's setting choice is $j$ and Bob's measurement outcome is $\gamma$ for the $k^{\rm th}$ pulse in an actual experiment. Importantly, \textcolor{black}{$q_{es}$ and $q_{obs_{j,\gamma}}$} are disjoint events in the entanglement-based protocol.

Next, for each of the setting choices $j$, we select \textcolor{black}{a} reference state $\ket{\phi_j}_{B}$ \textcolor{black}{for each} actual state $\ket{\psi_j}_{B}$. These reference states \textcolor{black}{belong to a certain} Hilbert space and by exploiting the properties of this Hilbert space, we may come up with a relationship for the $k^{\rm th}$ pulse (similar to Eq.~(39)) as
\begin{eqnarray}
0 \le f\left(P^{(k)}(q_{es}|{\rm Ref}), \textcolor{black}{\{P^{(k)}(q_{obs_{j,\gamma}}|{\rm Ref})\}_{j=1, 2,\cdots, m; \gamma=1,2,\cdots}}\right).
\label{mother-ineq-kth}
\end{eqnarray}
Note that, in Eq.~(\ref{mother-ineq-kth}), $P^{(k)}(q_{es}|{\rm Ref})$ and \textcolor{black}{$P^{(k)}(q_{obs_{j,\gamma}}|{\rm Ref})$} are the probabilities for the events $q_{es}$ and \textcolor{black}{$q_{obs_{j,\gamma}}$}, respectively, conditional on the selection of the reference states, rather than the actual states. We remark that, we have some freedom in selecting the function $f$ depending on \textcolor{black}{our} choice of the reference states. \textcolor{black}{However,} by taking into account the discussion in the previous section, it is preferable, \textcolor{black}{if possible,} to choose these states such that $f$ is loss-tolerant. 
%That is, when we regard the reference states as the actual states and, as a consequence, regard the probabilities in Eq.~(\ref{mother-ineq-kth}) as those in the actual protocol, no enhancement of $P^{(k)}(q_{es}|{\rm Ref})$ \textcolor{black}{should} occur with the increase of channel loss. 
In other words, the resulting secret key rate \textcolor{black}{should} be robust against \textcolor{black}{channel loss, especially in the presence of SPFs}. We emphasise, however, that if this loss-tolerant property is not \textcolor{black}{considered}, then the RT accepts a wide variety of functions as long as they can be related with the events corresponding to the actual states, after applying the Deviation evaluation part. In what follows, we consider \textcolor{black}{in more detail} the sufficient conditions to apply the RT, and we present how to execute the Deviation evaluation part depending on the function $f$.

\item \underline{Deviation evaluation part}

Before investigating the conditions for the function $f$, \textcolor{black}{we} first recall the key idea explained in the Main text. In general, this idea can be stated as follows: for any coefficient $c_{l_A}\ge0$ and probability $P(q_{l_A, l_B}|\xi):=p_{l_A}p_{l_B}{\rm Tr}\big[\dyad{\xi_{l_A}}{\xi_{l_A}}_B{\hat M}_{l_B}\big]$, \textcolor{black}{where} $p_{l_A}$ \textcolor{black}{is} the probability that Alice chooses the setting $l_A$ and $p_{l_B}$ is the probability 
%associated with the event $l_B$ on Bob's side, we have that
that Bob chooses the measurement basis associated with the event $l_B$ (here we define Bob's POVMs depending on Bob's basis choice separately, similarly to the Main text), we have that
%
%
%- Page 5, right after Eq. (19): I still find the statement "we have phi in the definition of P(qlA,lB|phi) in order to highlight the correspondence between the setting lA and the state phi" very confusing; moreover please note that "phi" is not a state, for that one should write \ket{phi}. If Kiyo has a suggestion for a better statement, please include it. Otherwise, I would simply delete that statement and do not say nothing about phi. But this is just a suggestion.
%
\begin{align}
\sum_{(l_A, l_B)\in \zeta} \textcolor{black}{c_{l_A}} P(q_{l_A, l_B}|\xi)= S\sum_{(l_A, l_B)\in \zeta} \frac{a_{{l_A, l_B}}}{S} {\rm Tr}\big[\dyad{\xi_{l_A}}{\xi_{l_A}}_B{\hat M}_{l_B}\big]=S {\rm Tr}\big[\dyad{\Omega}{\Omega}_{CB}{\hat M}],
\label{general-technique-prob}
\end{align}
with
\begin{align}
&\ket{\Omega}_{CB}:=\sum_{(l_A, l_B)\in \zeta}\sqrt{\frac{a_{l_A, l_B}}{S}}\textcolor{black}{\ket{l_A}_{C}}\ket{\xi_{l_A}}_{B},\nonumber\\
&{\hat M}:=\sum_{(l_A,\textcolor{black}{l_B})\in \zeta} \textcolor{black}{\dyad{l_A}{l_A}_C}\otimes {\hat M}_{l_B}.
\label{eq:meas}
\end{align}
\textcolor{black}{In  Eqs.~(\ref{general-technique-prob}) and (\ref{eq:meas}), 
%$c_{l_A}$ is a parameter related to the set of states $\{\ket{\phi_{l_A}}\}_{l_A}$, 
the coefficient $a_{l_A,l_B} := c_{l_A} p_{l_A} p_{l_B}$,} $S:=\sum_{(l_A, l_B)\in \zeta}\textcolor{black}{a_{l_A, l_B}}$ is the normalisation factor, \textcolor{black}{$\hat M_{l_B}$ is the POVM element corresponding to the event specified by $l_B$,} $\zeta$ is a subset of the set of disjoint events in the entanglement-based protocol, and \textcolor{black}{$\{\ket{l_A}_{C}\}_{l_A}$} is an orthonormal basis. Note that, $\xi$ can be any set of states, and in the context of the RT, it can be the actual states (Act) or the reference states (Ref). Similarly, $q_{l_A,l_B}$ can be any quantity, such as $q_{es}$ or $q_{obs_{j,\gamma}}$. Then, we can apply the bound defined in \textcolor{black}{Eq.~(\ref{eq:bound})} to the term $\Tr[\cdot]$ in Eq.~(\ref{general-technique-prob}) to evaluate the deviation of this probability from the probability associated with the actual states. Now, we consider the following three sufficient conditions for \textcolor{black}{the function} $f$.

\begin{enumerate}

\item \textit{When $f$ is a linear function and some of the coefficients in front of the probabilities associated \textcolor{black}{to the} reference states are negative, and the \textcolor{black}{others} are positive.}

This is the case evaluated in the Main text, where we simply divided the function $f$ into a positive part and a negative part to obtain
\begin{align}
&\textcolor{black}{0 \le} f\left(P^{(k)}(q_{es}|{\rm Ref}), \textcolor{black}{\{P^{(k)}(q_{obs_{j,\gamma}}|{\rm Ref})\}_{j=1, 2,\cdots, m; \gamma=1,2,\cdots}}\right) \nonumber \\
&~=S_{+} {\rm Tr}\big[\dyad{R_{+}}{R_{+}}_{CB}{\hat M}^{(k)}_{+}\big]-S_{-} {\rm Tr}\big[\dyad{R_{-}}{R_{-}}_{CB}{\hat M}^{(k)}_{-}\big],
\label{eq:(a)}
\end{align}
for appropriate choices of $\ket{R_+}_{CB}$, ${\hat M}_{+}^{(k)}$, $\ket{R_-}_{CB}$ and ${\hat M}_{-}^{(k)}$. Then, we can apply \textcolor{black}{Eq.~(\ref{eq:bound})} to each of the terms separately. \textcolor{black}{As} a result, we obtain
\begin{align} 
0&\le S_{+} \textcolor{black}{g^U} \left({\rm Tr}\big[\dyad{A_{+}}{A_{+}}_{CB}{\hat M}^{(k)}_{+}\big],  |{}_{CB}\langle A_+\ket{R_+}_{CB}|\right) \nonumber \\
&- S_{-} \textcolor{black}{g^L}\left({\rm Tr}\big[\dyad{A_{-}}{A_{-}}_{CB}{\hat M}^{(k)}_{-}\big],  |{}_{CB}\langle A_-\ket{R_-}_{CB}|\right),
\label{eq:(a)2}
\end{align}
where $\ket{A_+}_{CB}$ and $\ket{A_-}_{CB}$ are states associated with the actual states (see the Main text for specific examples). This means that, as required, the probability for the quantity $q_{es}$ is now expressed by using the actual states. \textcolor{black}{The transformation of the probabilities to the associated numbers shown in Eq.~(\ref{eq:(a)2}) can be done in exactly the same manner as in the Main text (see the discussions between \textcolor{black}{Eqs.~(53) and (55)}) because of the linearity of the function $f$.}

\item \textit{When $f$ is a linear function and all the coefficients in front of the probabilities associated to the reference states have the same sign.}

\textcolor{black}{This} is a special case of (a). A relationship for the actual states can be obtained by considering only the positive part or only the negative part of Eqs.~(\ref{eq:(a)}) and (\ref{eq:(a)2}).

\if0
For simplicity, we assume that the sign of the coefficients are positive, and the same argument holds for the negative sign. We have already seen this case in the security proof for the four-state protocol in Sec. \ref{GLLT-LT-as an application of RT}. Suppose that we have the reference formula (like the one in Eq. (\ref{four-state-virtual-state-expansion-method})), and we re-express the summed probabilities with the positive coefficients in the reference formula to obtain
\begin{eqnarray}
S_{+} {\rm Tr}[\ket{R_{+}}_{C, B}\bra{R_{+}}{\hat M}^{(k)}_{+}]
\end{eqnarray}
for appropriate choices of $\ket{R_{+}}_{C, B}$ and ${\hat M}^{(k)}_{+}$. Then, we apply Eq. (\ref{main-bound-method}) to obtain a relationship between the reference and actual states as
\begin{eqnarray}
{\rm Tr}[\ket{R_{+}}_{C, B}\bra{R_{+}}{\hat M}^{(k)}_{+}]=g({\rm Tr}\left[\ket{A_{+}}_{C, B}\bra{A_{+}}{\hat M}^{(k)}_{+}\right], |{}_{C, B}\langle A_{+}\ket{R_{+}}_{C, B}|, w)\,,
\end{eqnarray}
where $\ket{A_+}_{C, B}$ is a state associated to the actual states. Then, by plugging this relationship into the reference formula, we have 
\begin{eqnarray}
S_{+}g({\rm Tr}\left[\ket{A_{+}}_{C, B}\bra{A_{+}}{\hat M}^{(k)}_{+}\right], |{}_{C, B}\langle A_{+}\ket{R_{+}}_{C, B}|, 1)\,,
\end{eqnarray}
and we can continue the security proof with the same manner as before.
\fi

\item \textit{When $f$ is a monotone and concave function with respect to each of its arguments (in this case, an extra assumption is required, \textcolor{black}{see Eq.~(\ref{summed-monotone-concave-method}) below}).}

Here, we need to apply \textcolor{black}{Eq.~(\ref{eq:bound})} to each of its arguments separately. For instance, suppose that  
\begin{eqnarray}
\textcolor{black}{P^{(k)}(q_{obs_{j,\gamma}}|{\rm Ref})}=\textcolor{black}{p_{j,\gamma}}{\rm Tr}\big[\dyad{\phi_j}{\phi_j}_B {\hat M}_{\gamma}^{(k)}\big]\,,
\end{eqnarray}
where \textcolor{black}{$p_{j,\gamma}$} is a positive constant, which includes probabilities for the setting and measurement \textcolor{black}{basis} choices, $\ket{\phi_j}_{B}$ is the reference state for the $j^{\rm th}$ state and ${\hat M}_{\gamma}^{(k)}$ is a POVM element for Bob's outcome $\gamma$ after Eve's attack, conditional on the previous $k-1$ events. In this case, we employ \textcolor{black}{Eq.~(\ref{eq:bound})} and obtain
\textcolor{black}{
\begin{align}
&p_{j,\gamma} g^L \left(\frac{P^{(k)}(q_{obs_{j,\gamma}}|{\rm Act})}{p_{j,\gamma}'}, |{}_{B}\langle\psi_j\ket{\phi_j}_{B}|\right) \le P^{(k)}(q_{obs_{j,\gamma}}|{\rm Ref}) \nonumber \\
&\le p_{j,\gamma} g^U\left(\frac{P^{(k)}(q_{obs_{j,\gamma}}|{\rm Act})}{p_{j,\gamma}'}, |{}_{B}\langle\psi_j\ket{\phi_j}_{B}|\right).
\end{align}}
Here, \textcolor{black}{$p_{j,\gamma}'$} corresponds to \textcolor{black}{$p_{j,\gamma}$} after some modification. Similarly, we have that 
\textcolor{black}{
\begin{align}
{p^{es}} g^L \left(\frac{P^{(k)}(q_{es}|{\rm Act})}{p^{es'}}, |{}_{B}\langle\psi_{es}\ket{\phi_{es}}_{B}| \right) \le {P^{(k)}(q_{es}|{\rm Ref})} \le {p^{es}} g^U \left(\frac{P^{(k)}(q_{es}|{\rm Act})}{p^{es'}}, |{}_{B}\langle\psi_{es}\ket{\phi_{es}}_{B}| \right)\,,
\label{eq:es_ref}
\end{align}}where $p^{es}$ ($p^{es'}$) is the probability associated to the selection of the quantity that needs to be estimated in the security proof, when employing the reference (actual) states. Note that, in general we \textcolor{black}{have the freedom to} choose $p^{es'}$, however, it is likely that the choice of $p^{es'}=p^{es}$ results in a higher secret key rate. Moreover, \textcolor{black}{in Eq.~(\ref{eq:es_ref})} $\ket{\phi_{es}}_{B}$ ($\ket{\psi_{es}}_{B}$) \textcolor{black}{denotes} the reference (actual) state associated with the quantity to be estimated. Therefore, we obtain
%\begin{align}
%0&\le f\Bigg(\textcolor{black}{{p^{es}}} g\left(\frac{P^{(k)}(q_{es}|{\rm Act})}{p^{es'}}, |{}_{B}\langle\psi_{es}\ket{\phi_{es}}_{B}|, \omega_{es}\right), \nonumber \\
%&~~~~~ \left \{ \textcolor{black}{p_{j,\gamma}} g\left(\textcolor{black}{\frac{P^{(k)}(q_{obs_{j,\gamma}}|{\rm Act})}{p_{j,\gamma}'}}, |{}_{B}\langle\psi_j\ket{\phi_j}_{B}|, \textcolor{black}{\omega_{j,\gamma}}\right) \right \}_{\textcolor{black}{j=1, 2,\cdots, m; \gamma=1,2,\cdots}}\Bigg)\nonumber\\
%&\le f\Bigg(\textcolor{black}{{p^{es}}} g\left(\frac{P^{(k)}(q_{es}|{\rm Act})}{p^{es'}}, |{}_{B}\langle\psi_{es}\ket{\phi_{es}}_{B}|, \omega_{es}^{*}\right), \nonumber \\ 
%&~~~~~ \left \{\textcolor{black}{p_{j,\gamma}} g\left(\textcolor{black}{\frac{P^{(k)}(q_{obs_{j,\gamma}}|{\rm Act})}{p_{j,\gamma}'}}, |{}_{B}\langle\psi_j\ket{\phi_j}_{B}|, \omega_{j,\gamma}^{*}\right) \right \}_{\textcolor{black}{j=1, 2,\cdots, m; \gamma=1,2,\cdots}}\Bigg) 
%\end{align}
\textcolor{black}{
\begin{align}
0 &\le f \Big(P^{(k)} (q_{es}|{\rm Ref}), \{P^{(k)} (q_{obs_{j,\gamma}}|{\rm Ref})\}_{j=1, 2,\cdots, m; \gamma=1,2,\cdots}\Big) \nonumber \\
&\le f\Bigg(\textcolor{black}{{p^{es}}} \textcolor{black}{g^*} \left(\frac{P^{(k)}(q_{es}|{\rm Act})}{p^{es'}}, |{}_{B}\langle\psi_{es}\ket{\phi_{es}}_{B}| \right), \nonumber \\ 
&~~~~~ \left \{\textcolor{black}{p_{j,\gamma}} \textcolor{black}{g^*}\left(\textcolor{black}{\frac{P^{(k)}(q_{obs_{j,\gamma}}|{\rm Act})}{p_{j,\gamma}'}}, |{}_{B}\langle\psi_j\ket{\phi_j}_{B}| \right) \right \}_{\textcolor{black}{j=1, 2,\cdots, m; \gamma=1,2,\cdots}}\Bigg) 
\label{monotone-concave-method}
\end{align}}where in the second inequality, we have used the monotonicity of the function $f$ to maximise the function $f$ with respect to \textcolor{black}{$P^{(k)} (q_{es}|{\rm Ref})$ and $P^{(k)} (q_{obs_{j,\gamma}}|{\rm Ref})$,} and we denote the optimal values as \textcolor{black}{$g^*(x,y)$,} i.e.~each of them is either \textcolor{black}{$g^L(x,y)$ or $g^U(x,y)$.} 
\end{enumerate}
Note that, in cases (a), (b) and (c) a relationship for the actual states is obtained, and this finishes the Deviation evaluation part of the RT.
\end{enumerate}

Finally, as explained in the Main text, we have to convert the inequality obtained in the Deviation evaluation part into a relationship in terms of numbers, rather than probabilities. For this, we first take a summation over $k\in\{1, 2, \cdots, N\}$. Using case (c) for generality, we have that
\textcolor{black}{
\begin{align}
0&\le\frac{1}{N}\sum_{k=1}^{N}f\Bigg(\textcolor{black}{{p^{es}}} \textcolor{black}{g^*} \left(\frac{P^{(k)}(q_{es}|{\rm Act})}{p^{es'}}, |{}_{B}\langle\psi_{es}\ket{\phi_{es}}_{B}| \right), \nonumber \\ 
&~~~~~~~~~~~~~~\left \{\textcolor{black}{p_{j,\gamma}} \textcolor{black}{g^*}\left(\textcolor{black}{\frac{P^{(k)}(q_{obs_{j,\gamma}}|{\rm Act})}{p_{j,\gamma}'}}, |{}_{B}\langle\psi_j\ket{\phi_j}_{B}|\right)\right \}_{\textcolor{black}{j=1, 2,\cdots, m; \gamma=1,2,\cdots}}\Bigg)\nonumber\\
&\le f\Bigg(\sum_{\textcolor{black}{k=1}}^{N}\frac{1}{N}\textcolor{black}{{p^{es}}} \textcolor{black}{g^*} \left(\frac{P^{(k)}(q_{es}|{\rm Act})}{p^{es'}}, |{}_{B}\langle\psi_{es}\ket{\phi_{es}}_{B}|\right),\nonumber \\
&~~~~~~\left \{\sum_{\textcolor{black}{k=1}}^{N}\frac{1}{N} \textcolor{black}{p_{j,\gamma}} \textcolor{black}{g^*}\left(\textcolor{black}{\frac{P^{(k)}(q_{obs_{j,\gamma}}|{\rm Act})}{p_{j,\gamma}'}}, |{}_{B}\langle\psi_j\ket{\phi_j}_{B}| \right) \right \}_{\textcolor{black}{j=1, 2,\cdots, m; \gamma=1,2,\cdots}}\Bigg)\nonumber\\
&\le f\Bigg(\textcolor{black}{{p^{es}}} \textcolor{black}{g^*}\left(\sum_{\textcolor{black}{k=1}}^{N}\frac{P^{(k)}(q_{es}|{\rm Act})}{Np^{es'}}, |{}_{B}\langle\psi_{es}\ket{\phi_{es}}_{B}|\right), \nonumber \\
&~~~~~~ \left \{\textcolor{black}{p_{j,\gamma}} \textcolor{black}{g^*}\left(\sum_{\textcolor{black}{k=1}}^{N}\textcolor{black}{\frac{P^{(k)}(q_{obs_{j,\gamma}}|{\rm Act})}{Np_{j,\gamma}'}}, |{}_{B}\langle\psi_j\ket{\phi_j}_{B}|\right) \right \}_{\textcolor{black}{j=1, 2,\cdots, m; \gamma=1,2,\cdots}}\Bigg),
\label{summed-monotone-concave-method}
\end{align}}where in the second inequality, we exploit the concavity of the function $f$ with respect to each of \textcolor{black}{its} arguments, i.e.~$\sum_{k} f(\{x_{k,i}\}_{i})/N\le f(\{\sum_{k} x_{k,i}\}_{i}/N)$ \textcolor{black}{with $N$ being a positive integer}. Importantly, the relationship through the third inequality is the extra sufficient condition that is required to apply the RT \textcolor{black}{to the case (c)}. \textcolor{black}{This extra assumption is required here because we need a relationship in terms of numbers in the security proof. Note that, in cases \textcolor{black}{(a) and (b)} the function $f$ is linear and thus Eq.~(\ref{eq:(a)2}) can be readily transformed into an expression in terms of numbers.} Then, as before we apply Azuma's inequality [37] or Kato's inequality [38] to each of the sums. As a result, we have that in the asymptotic limit of large $N$ 
\textcolor{black}{
\begin{align}
0\le ~&f\Bigg(\textcolor{black}{{p^{es}}} \textcolor{black}{g^*}\left(\frac{N(q_{es}|{\rm Act})}{Np^{es'}}, |{}_{B}\langle\psi_{es}\ket{\phi_{es}}_{B}|\right), \nonumber \\
&~~ \left \{\textcolor{black}{p_{j,\gamma}} \textcolor{black}{g^*}\left(\textcolor{black}{\frac{N(q_{obs_{j,\gamma}}|{\rm Act})}{Np_{j,\gamma}'}}, |{}_{B}\langle\psi_j\ket{\phi_j}_{B}| \right) \right \}_{\textcolor{black}{j=1, 2,\cdots, m; \gamma=1,2,\cdots}}\Bigg),
\label{summed-monotone-concave-method2}
\end{align}}where $N(q_{es}|{\rm Act})$ and \textcolor{black}{$N(q_{obs_{j,\gamma}}|{\rm Act})$} represent the actual number of the occurrences \textcolor{black}{associated} with the events $q_{es}$ and \textcolor{black}{$q_{obs_{j,\gamma}}$}, respectively, in the actual protocol.

%%%%%%%%%%%%%%%%%

\end{document}